\documentclass[aps,jcp,superscriptaddress,amsmath,amssymb]{revtex4-1}

\usepackage{comment}
\usepackage{tikz}
\usetikzlibrary{trees}
\usetikzlibrary{decorations.pathmorphing}
\usetikzlibrary{decorations.markings}
\usepackage{tikz-cd}
\usetikzlibrary{arrows,arrows.meta}
\usetikzlibrary{decorations.pathreplacing,snakes}
\tikzset{
    photon/.style={decorate, decoration={snake,segment length=1.5mm}, draw=black},
    coulomb/.style={dotted},
    electron/.style={draw=black, postaction={decorate},
        decoration={markings,mark=at position .55 with {\arrow[draw=black]{>}}}}, 
    gluon/.style={decorate, draw=magenta,
        decoration={coil,amplitude=4pt, segment length=5pt}},
    boundelectron/.style={thick, double},
    transverse/.style={dashed},
    marrow/.style={decoration={markings,mark=at position 0.5 with {\arrow{#1}}}, postaction=decorate}    
}
\usepackage{graphicx}% Include figure files
\usepackage{dcolumn}% Align table columns on decimal point
\usepackage{bm}% bold math
\usepackage{rotating}

\newcolumntype{.}{D{.}{.}{8}}

\usepackage[utf8]{inputenc}
\usepackage{physics}
\usepackage{amsmath, amssymb}
\usepackage{tabularx,booktabs,array,dcolumn}
\usepackage{setspace}
\usepackage{vmargin}
\usepackage{xcolor}
\usepackage{graphicx,psfrag,subfigure}

\usepackage{float}
\usepackage[all]{xy}

\usepackage{bm}
\usepackage{mathtools}
\usepackage{physics}
\usepackage[english]{babel}

\newcommand{\bos}[1]{\boldsymbol{#1}}
\newcommand{\mr}[1]{\mathrm{#1}}
\newcommand{\pd}[2]{\frac{\partial #1}{\partial #2}}

\def\Eh{E_\mathrm{h}}

\def\iim{\mr{i}}
\def\eem{\mr{e}}

\def\nb{N_\text{b}} %no. of spatial basis functions

\def\unitfour{1^{[4]}} %{1_4}
\def\four{^{[4]}} %{^{[2]}}
\def\ta{\tilde{\alpha}}

\def\dd{\mathrm{d}}

\def\bp{\bos{p}}
\def\br{\bos{r}}
\def\bx{\bos{r}}

\def\bk{\bos{k}}
\def\balpha{\bos{\alpha}}
\def\bsigma{\bos{\sigma}}

\def\epsi{\varepsilon}
\def\mcE{\mathcal{E}}
\def\mcF{\mathcal{F}}
\def\mcG{\mathcal{K}}%{\mathcal{G}}
\def\mcH{\mathcal{H}}
\def\mcD{\mathcal{D}}
\def\mcL{\mathcal{L}}

\def\mcV{\mathcal{U}}
\def\mcD{\mathcal{D}}

\def\pupu{\scalebox{0.65}{$++$}}
\def\pumi{\scalebox{0.65}{$+-$}}
\def\mipu{\scalebox{0.65}{$-+$}}
\def\mimi{\scalebox{0.65}{$--$}}
\def\plusz{{\scriptsize +}}

\def\balpha{\boldsymbol{\alpha}}

\def\rad{\text{rad}}
\def\col{\text{C}}
\def\tra{\text{T}}
\def\Breit{\text{B}}
\def\tI{\text{I}}
\def\tC{\text{C}}
\def\tB{\text{B}}
\def\tCB{\text{CB}}
\def\tT{\text{T}}

\def\tc{\text{i}}

\def\Ic{\mathcal{I}_\text{i}}
\def\phic{\Phi_\text{i}}
\def\Ec{E_\text{i}}
\def\kernel{K}
\def\smallkernel{\kappa}
\def\green{G}
\def\Konevp{{\green'_1}}%{{K'_{1V}}}
\def\Konev{\green_1}%{K_{1V}}
\def\Ktwovp{{\green'_2}}%{{K'_{2V}}}
\def\Ktwov{{\green_2}}%{K_{2V}}
\def\Sigonev{\Sigma_1}%{\Sigma_{1V}}
\def\Sigtwov{\Sigma_2}%{\Sigma_{2V}}
\def\barG{\kernel} %{\bar{G}}
\def\Df{D_\text{F}}

\def\Psitot{\Psi_\text{T}}
\def\kernelrelt{\kernel_\text{t}}
\def\Psimom{\bos{\psi}}

\def\nulla{{(0)}}

\usepackage[unicode]{hyperref}
\usepackage{soul}

\usepackage[unicode]{hyperref}
\hypersetup{
   unicode=true,          % non-Latin characters in Acrobat??s bookmarks
   plainpages=false,
   colorlinks=true,       % false: boxed links; true: colored links
   linkcolor=blue,          % color of internal links
   citecolor=blue,        % color of links to bibliography
}

\usepackage{url}

\begin{document}

\title{%
The Bethe--Salpeter QED wave equation for bound-state computations of atoms and molecules
}

\author{Edit M\'atyus} 
\email{edit.matyus@ttk.elte.hu}
\author{D\'avid Ferenc} 
\author{P\'eter Jeszenszki} 
\author{\'Ad\'am Marg\'ocsy}
\affiliation{ELTE, Eötvös Loránd University, Institute of Chemistry, 
Pázmány Péter sétány 1/A, Budapest, H-1117, Hungary}

\date{\today}

\begin{abstract}
\noindent %
Interactions in atomic and molecular systems are dominated by electromagnetic forces and the theoretical framework must be in the quantum regime. 
The physical theory for the combination of quantum mechanics and electromagnetism, quantum electrodynamics has been `established' by the mid-twentieth century, primarily as a scattering theory. 
To describe atoms and molecules, it is important to consider bound states. 
In the non-relativistic quantum mechanics framework, bound states can be efficiently computed using robust and general methodologies with systematic approximations developed for solving \emph{wave equations.}
With the sight of the development of a computational quantum electrodynamics framework for atomic and molecular matter, the field theoretic Bethe--Salpeter wave equation expressed in space-time coordinates, its exact equal-time variant and emergence of a relativistic wave equation is reviewed. 
A computational framework, with initial applications and future challenges in relation with precision spectroscopy, is also highlighted.\\[1cm]
\end{abstract}

\maketitle

%\input{ms}
%%%%%%%%%%%%%%%%%%%%%%%%%%%%%%%%%%%%%%%%%%%%%%%%%%%%%%%%
\section{Introduction: a historical line-up \label{sec:intro}}
\noindent%
Dirac's one-electron space-time equation was an ingenious departure from Schrödinger's time-dependent wave equation to have a Lorentz covariant description, but it was strange due to the introduction of hole theory that seemed a bit artificial \cite{Di28a,Di28b}. 
In a recorded discussion from 1982, Dirac modestly admitted to Hund that for him, it took a year, perhaps two, to understand the role of the negative-energy states \cite{DiHu82}. 

Breit attempted a two-particle generalization of Dirac's one-electron theory in a series of papers between 1928 and 1931 \cite{Br28,Br29,Br30,Br31}, by adopting Darwin's 1920 calculation of the classical electromagnetic interaction for two moving charges  \cite{Da20} and the quantum mechanical velocity operator obtained with Dirac's formalism \cite{Br31}. Already from the beginning, it was apparent that the Breit equation was not Lorentz covariant. Nevertheless,  Breit used this `quasi-relativistic' equation in a perturbation theory approach imposing the Pauli approximation to the four-(sixteen-)component wave function. Good agreement with experiment was obtained after discarding a term from the result `by hand'~\cite{Br29,Br30}. This procedure was later explained based on Dirac's hole theory by Brown and Ravenhall \cite{BrRa51}.

The problem, called Brown--Ravenhall (BR) disease, related to the artificial coupling of the positive- and negative-energy states of Dirac's theory when na\"ively applied to two-particle systems, survived also in the modern literature and it is commonly used to explain the failure of the two-(many-)particle Breit equation. A recent numerical study demonstrated that bound states of helium-like two-electron systems represented by the Breit equation have (unphysical) finite lifetimes (on order $\alpha^3\Eh$, where $\alpha$ is the fine-structure constant) \cite{PeByKa06,PeByKa07}. 
It has been also discussed~\cite{Su83} that there was no BR dissolution problem for isolated two-particle systems, like positronium, when modelled with the Breit equation, and it was numerically demonstrated by a finite-element computation \cite{ScShMo92} that the energy levels of the two-particle Breit equation (with Coulomb interactions, no external fields) are `stable'. Nevertheless, the two-particle Breit equation is still incorrect, or in other words, `correct only up to order $\alpha^2\Eh$' \cite{PeByKa07}. 

A consistent and Lorentz covariant many-particle theory was put forward by the development of quantum electrodynamics (QED).
As a natural continuation of Feynman's two papers in 1949 on the reinterpretation of the mathematical solutions of the Dirac equation~\cite{Fe49a} and the development of the space-time approach to quantum electrodynamics~\cite{Fe49b}, Salpeter and Bethe in 1951 published a Lorentz-covariant wave equation for two interacting particles (with stating that generalization to more than two particles is straightforward) \cite{SaBe51}. It is interesting to note that the same equation was written at the end of a paper without explanation by Nambu already in 1950~\cite{Na50}, and it was formulated by Schwinger~\cite{Sc51} and also by Gell-Mann and Low~\cite{GMLo51} during 1951.
In 1952, Salpeter used this equation for the hydrogen atom in combination with perturbation theory and an instantaneous interaction kernel. Probably, this was the first formulation of the exact equal-time equation for two-particle systems~\cite{Sa52}.  Salpeter reported results for the hydrogenic case (of one heavy and one light particle, $M\gg m$) up to the $\alpha^3(m/M)\Eh$ order, and he stated that the calculation can be generalized to any masses. In 1954, Fulton and Martin calculated the energy levels for an arbitrary two-fermion system, such as positronium, up to $\alpha^3\Eh$ order \cite{FuMa54}.

In 1958, Sucher's PhD thesis represented another important step forward using the formalism and extending Salpeter's work to a two-electron system in an external Coulomb field, for the example of the helium atom \cite{sucherPhD1958}. Sucher's final $\alpha^3\Eh$-order correction formulae were identical with those reported by Araki \cite{araki57} a year earlier, but we can build on the fundamental ideas explained in Sucher's work for further developments.

In 1974, Douglas and Kroll \cite{DoKr74} started their paper on the $\alpha^4\Eh$-order corrections to the fine-structure splitting of helium with a good review of Sucher's work by extending the formalism with explicitly writing also the radiative terms in the wave equation (Sucher only highlighted the steps at the end of his work). 

Then, in 1989, Adkins elaborated this direction for positronium, still relying on a perturbative expansion with respect to the non-relativistic reference for practical calculations \cite{Ad89}. During the 1990s, Zhang worked on higher-order corrections to the fine-structure splitting ($\alpha^5\Eh$) and energy levels ($\alpha^4\Eh$) of helium.

Pachucki initiated a different approach starting from the late 1990s \cite{Pa97-1,Pa97-2,Pa98-1,Pa98-2,Pa05FW}.
This approach is based on performing a Foldy--Wouthuysen transformation \cite{FoWo50} of the Dirac operator in the Langrangian density---thereby linking the formalism to the non-relativistic theory from the outset, and then, collecting corrections to the poles of the equal-time Green function~\cite{Sh02} to the required $\alpha$ order. In 2006, Pachucki reported the complete $\alpha^4\Eh$-order corrections to the energy levels of singlet helium \cite{Pa06}, thereby extending the 1974 work of Douglas and Kroll valid only for triplet states, as well as work from Yelkhovsky \cite{Ye01} and computations from Korobov and Yelkhovsky \cite{KoYe01} in 2001 for $\alpha^4\Eh$-order corrections of singlet helium.
In 2016, the complete $\alpha^4\Eh$-order corrections derived by Pachucki were used for the ground electronic state of the H$_2$ molecule with fixed protons~\cite{PuKoCzPa16}. 
Most recently, the `Foldy--Wouthuysen--Pachucki' approach has been used to derive $\alpha^5\Eh$-order contributions for triplet states of helium \cite{PaYePa20,PaYePa21}. 

In contrast to using a non-relativistic reference (as in all previous work), we aim for a \emph{relativistic} QED approach, in which some (well-defined, many-particle) relativistic wave equation is first solved to high precision and used as a reference for computing `QED' (retardation, pair, and radiative) corrections up to a required accuracy. 
Such an approach appears to be feasible along the lines formally started by Bethe, Salpeter, Sucher, Douglas and Kroll. These authors performed calculations by hand, so in the end, they had to rely on approximations based on the non-relativistic formalism.
Nowadays, we can use the power of modern computers to first numerically solve a many-particle relativistic wave equation, and then, compute corrections to the relativistic energy. 
It is also necessary to add that there have been several articles on understanding and solving the original, space-time Bethe--Salpeter (BS) equation \cite{Wi54,Cu54,Sc65,Na69,La87}. 
For atomic and molecular computations, the exact equal-time form of the BS equation, as introduced by Salpeter and Sucher \cite{Sa52,sucherPhD1958}, appears to be more promising. In this approach, a \emph{two-particle} relativistic Hamiltonian and corresponding wave equation emerges, for which numerical strategies for solving wave equations, including the variational method, can be used. 
During the 1980s, Sucher \cite{Su80,Su83,Su84} published review articles about the (formal) connection of the equal-time BS wave equation with the relativistic quantum chemistry framework and computational methodologies, 
\emph{e.g.,} \cite{dirac2020,BeDSQuTaSt20}. 
In this context, it is necessary to mention the excellent book of Lindgren who further developed these ideas for orbital-based many-body applications in chemistry \cite{lindgrenRelativisticManyBodyTheory2011}.

We consider the renormalized, `mixed gauge', two-particle Bethe--Salpeter equation as the starting point for a theoretical framework of atoms and molecules and with relevance for spectroscopic applications. 
This theoretical framework is reviewed in the first part of the paper by relying on work by Sucher \cite{sucherPhD1958}, Douglas and Kroll \cite{DoKr74}, as well as Salpeter \cite{Sa52}.
The second part of the article highlights our recent work \cite{JeFeMa21,JeFeMa22,FeJeMa22,FeJeMa22b,JeMa22,FeMa22Ps}, algorithmic details of a computer implementation and numerical results for two spin-1/2 particles with and without a fixed, external Coulomb field, \emph{i.e.,} with relevance for relativistic Born--Oppenheimer (BO) as well as relativistic pre-Born--Oppenheimer (pre-BO) computations. 
Although in the present review, we focus on the theory and a numerical procedure for two-particle systems, we mention 
Sucher's series of papers \cite{Su80,Su83,Su84} from the 1980s implying a possible generalization and 
Broyles' work from 1987 \cite{Br87} about presenting a line of thoughts connecting field theory and an $N$-particle no-pair Dirac--Coulomb--Breit wave equation.

Regarding a relativistic QED approach, we also mention the quasi-potential method, which originates from Logunov, Tavkhelidze, and Faustov working during the 1960-70s \cite{LoTa63,Fa70}, and the corresponding two-time (equal-time) Green function idea developed by Shabaev \cite{Sh02}. 
Comparison of the Salpeter--Sucher approach with the quasi-potential method is left for future work.

%%%%%%%%%%%%%%%%%%%%%%%%%%%%%%%%%%%%%%%%%%%%%%%%%%%%%%%%%%%%%%%%%%%%%%%%%%%%%%%%%%%%%%
%%%%%%%%%%%%%%%%%%%%%%%%%%%%%%%%%%%%%%%%%%%%%%%%%%%%%%%%%%%%%%%%%%%%%%%%%%%%%%%%%%%%%%
\section{The Bethe--Salpeter equation and the Salpeter--Sucher exact equal-time approach \label{sec:BSeq}}
\subsection{Introductory ideas and propagators}
The Dirac equation for a particle of mass $m_1$ and $x_1=(\bos{r}_1,t_1)$  space-time coordinates is 
\begin{align}
  \left[%
    \iim \pd{}{t_1} - H^\nulla_1
  \right] 
  \varphi_n^{(1)}(x_1) = 0 \; 
\end{align}
with the free-particle Hamiltonian 
\begin{align}
  H^\nulla_1=
  -\iim \bos{\alpha}_1 \bos{\nabla}_{1}  
  +
  \beta_1 m_1 \; 
\end{align}
and the $\bos{\alpha}_1$ and $\beta_1$ Dirac matrices.
Feynman pointed out in 1949 \cite{Fe49a} that instead of working with the Hamiltonian equation, it is often more convenient to use the corresponding Green function or propagator 
\begin{align}
  \left[%
    \iim \pd{}{t_1} - H^\nulla_1
  \right] 
  G_1^\nulla(x_1,x_1') = \iim \beta_1 \delta(x_1-x_1') \; .
  \label{eq:freepart}
\end{align}
For a Dirac particle in an external scalar field, 
$\Phi_1$
\begin{align}
  \left[%
    \iim \pd{}{t_1} - H_1
  \right]
  G_1(x_1,x_1') 
  = 
  \iim\beta_1\delta(x_1-x_1') \; 
  \label{eq:Hextfield}
\end{align}
with 
\begin{align}
  H_1 
  =
  H^\nulla_1+z_1e\Phi_1
  =
  H^\nulla_1+U_1 \; ,
  \label{eq:oneDirac}
\end{align}
$z_1\in\mathbb{Z}$ stands for the charge number of the active particle and $e$ is the elementary charge.
Simple calculation \cite{Fe49a} shows that 
the $G_1$ propagator can be obtained from the $G_1^{(0)}$ free-particle propagator through the integral equation (corresponding to subsequent interaction events of the particle with the external field) as
\begin{align}
  G_1(x_1,x_1')
  = 
  G_1^{(0)}(x_1,x_1')
  -\iim \int 
    G_1^{(0)}(x_1,y_1) \beta_1U_1(y_1) G_1(y_1,x_1')\ \dd y_1 \; .
\end{align}
According to Feynman's combination of the electronic and positronic theory in a consistent manner \cite{Fe49a}, the propagator is expressed with the eigenvalues and eigenfunctions of the Dirac Hamiltonian as the sum over positive-energy (electronic) states
%for
moving forward in time, and the negative sum over negative-energy (positronic) states
% for
moving backward in time. Feynman defined the free-particle propagator this way corresponding to Eq.~\eqref{eq:freepart}. 
The arguments can be taken over for a particle in an external field, Eq.~\eqref{eq:Hextfield}, which is known as the `Furry picture' \cite{Fu51}, and the propagator is
\begin{align}
  \Konev(x_1,x'_1)
  =
  \left\lbrace 
  \begin{array}{@{}rl@{}}
     \sum\limits_{E^{(1)}_n>0} 
       \phi^{(1)}_n(\bx_1) \bar{\phi}^{(1)}_n(\bx'_1) 
       \eem^{-\iim E^{(1)}_n(t_1-t'_1)}\; , & t_1>t'_1  \\
    -\sum\limits_{E^{(1)}_n<0} 
      \phi^{(1)}_n(\bx_1) \bar{\phi}^{(1)}_n(\bx'_1) \eem^{-\iim E^{(1)}_n(t_1-t'_1)}\; , & t_1<t'_1  \\    
  \end{array}
  \right. \; , 
\end{align}
where $\bar{\phi}_n^{(1)}=\phi_n^{(1)\ast} \beta_1$ is the Dirac adjoint.
The $\Konev^{(0)}(x_1,x'_1)$ free-particle propagator is recovered for eigenvalues and eigenfunctions of the Dirac equation with $\Phi_1=0$. 
Regarding the external field in the present work, only the scalar potential due to the Coulomb field of the fixed nuclei will be relevant, \emph{e.g.,} for helium-like systems with the nucleus fixed at the origin and with $Z$ nuclear charge number, the interaction energy is
\begin{align}
   U_1(\bx_1) = z_1 \frac{Z \alpha}{|\br_1|} \; ,
   \label{eq:nucleusfield}
\end{align}
where $\alpha=e^2/(4\pi)$ is the fine-structure constant
in natural units ($\hbar=c=\epsilon_0=1$).

To describe a two-particle system, we can consider the
$G(x_1,x_2;x_1',x_2')$ two-particle propagator or amplitude which describes that particles 1 and 2 get from $x_1',x_2'$ to $x_1,x_2$ space-time points. For non-interacting particles, the two-particle propagator is the simple product of the one-particle propagators, $G_1(x_1,x_1')G_2(x_2,x_2')$.
For interacting two-particle systems Salpeter and Bethe \cite{SaBe51}, following Feynman \cite{Fe49a,Fe49b}, devised an integral equation, called Bethe--Salpeter (BS) equation,
\begin{align}
  &\green(x_1,x_2;x'_1,x'_2) 
  =
  \Konev(x_1,x'_1) \Ktwov(x_2,x'_2) \nonumber\\
  &\quad\quad\quad- 
  \iim \int \dd y_1 \dd y_2 \dd y'_1 \dd y'_2\ 
  \Konev(x_1,y_1) \Ktwov(x_2,y_2) \barG(y_1,y_2;y'_1,y'_2) \green(y'_1,y'_2;x'_1,x'_2) \; ,
  \label{eq:BSint}
\end{align}
where $\barG$ is the interaction function. 
In particular, $\barG$ must contain only the so-called `irreducible' interactions, since the corresponding consecutive, so called `reducible', interactions are already included, `iterated' to all orders by the integral equation.

The simplest interaction function, $K^{(1)}$, corresponds to the single photon exchange (see also Sec.~\ref{sec:kernel}) 
with $\gamma_{i}^{\mu}=\beta_i(\bos{\alpha}_i,1)$ 
\begin{align}
  \barG^{(1)}(x_1,x_2;x'_1,x'_2)
  =
  \alpha
  z_1z_2
  \gamma_{1\mu} \gamma_{2\nu} \Df^{\mu\nu}[(x_1-x_2)^2] \delta^4(x_1-x'_1) \delta^4(x_2-x'_2) \; ,
  \label{eq:onephoton}
\end{align}
where $\Df^{\mu\nu}$ is the photon propagator, which takes a simple, manifestly covariant form in Feynman gauge,
\begin{align}
  \Df^{\mu\nu}[(x_1-x_2)^2]
  =
  -
 % \iim
  \int \frac{\dd^4 k}{(2\pi)^4}\ \frac{g^{\mu\nu}}{k^2+\iim \Delta}\  \eem^{-\iim k\cdot (x_{1}-x_2)} \; .
  \label{eq:photonFeynman}
\end{align}
To describe the inter-particle interaction in atoms and molecules, it is more convenient to use the Coulomb gauge,
in which the interaction is the sum of the Coulomb (C, the dominant part) and the transverse (T) contributions,
\begin{align}
  \barG^{(1)}
  (x_1,x_2;x'_1,x'_2)
  =
  [\barG^{(1)}_\tC(x_1,x_2) + \barG^{(1)}_\tT(x_1,x_2)]
  \delta(x_1-x'_1)\delta(x_2-x'_2)
  \label{eq:onephotCoul}
\end{align}
with
\begin{align}
  \barG^{(1)}_\tC(x_1,x_2)
  &=
  \alpha z_1z_2
  \beta_1\beta_2
  \int \frac{\dd^4k}{(2\pi)^4}
  \frac{4\pi}{\bk^2}\eem^{-\iim k (x_1-x_2)} 
  =\beta_1\beta_2
   \frac{\alpha z_1z_2}{|\bx_1-\bx_2|}
   \delta(t_1-t_2)
  \label{eq:Coulomb}
  \\
  \barG^{(1)}_\tT(x_1,x_2)
  &=
  \alpha z_1z_2
  \beta_1\beta_2  
  \int \frac{\dd^4 k}{(2\pi)^4}
  \frac{4\pi\ta_1^i \ta_2^i}{\omega^2-\bk^2+\iim \Delta } 
  \eem^{-\iim k (x_1-x_2)}
     \label{eq:transverse}  
  \; ,
\end{align} 
where
\begin{align}
 \tilde{\alpha}^i(\bk)=\left(\delta^{ij}-\frac{k^ik^j}{\bk^2}\right)\alpha^j
 \label{eq:aatransverse}
\end{align}
corresponds to the transverse components of $\boldsymbol{\alpha}$
perpendicular to $\bk$. (In Eq.~\eqref{eq:Coulomb}, we highlighted the well-known coordinate-space form of the Coulomb interaction, the primarily important interaction term in quantum chemistry).

If radiative corrections are accounted for, 
the one-electron (one-particle) propagator is replaced by \cite{SaBe51,Dy49,sucherPhD1958,DoKr74}
\begin{align}
  \Konevp
  =
  \Konev 
  + \Konev\Sigonev\Konev
  + \Konev\Sigonev\Konev\Sigonev\Konev + \ldots 
  =\Konev+\Konevp\Sigonev\Konev
  \; ,
  \label{eq:propagrad}
\end{align}
or equivalently,
\begin{align}
  (\Konevp)^{-1} 
  = 
  \Konev^{-1} - \Sigonev \; ,   
  \label{eq:invpropagrad}  
\end{align}
where $\Sigonev$ is the sum of the one-electron self-energy contributions. 
Douglas and Kroll \cite{DoKr74}, following the last chapter of Sucher's work \cite{sucherPhD1958}, formulated the two-electron equations by formally including all radiative corrections.
This formulation also relies on the work of Mathews and Salam \cite{MaSa54} who explained that the Bethe--Salpeter equation can be renormalized with the replacement of
$\Konevp$, $\Ktwovp$, $\gamma_{1 \mu}$, $\gamma_{2 \mu}$, and the $\Df$ photon propagator by \cite{sucherPhD1958,DoKr74}
\begin{align}
  \Konevp^\ast 
  &= 
  \Konev + \Konev \Sigonev^\ast \Konev + \ldots \; ,\\
  \Gamma_{1 \mu}^\ast 
  &= 
  \gamma_{1 \mu} + \Lambda_{1 \mu}^\ast \; , \\
  \Df^\ast 
  &= 
  \Df + \Df \Pi^\ast \Df + \ldots \; .
\end{align}
We note that Mathews and Salam mostly formulated their renormalization approach based on series expansion, whereas K\"allen \cite{Ka52} and Lehmann \cite{Le54} defined renormalization terms without the use of power series expansion in the interaction constant.
Karplus and Kroll \cite{KaKr50} and Jauch and Rohrlich \cite{JaRo76} carried out explicit calculations 
for the $\Sigonev^\ast$, $\Pi^\ast$, $\Lambda_{1 \mu}^\ast$ renormalized electron self-energy, photon self-energy, and vertex correction operators to order $\alpha$ for the case of no external potentials.

For renormalization, it is necessary to work in the Feynman gauge, Eq.~(\ref{eq:photonFeynman}). 
At the same time, binding of the particles in atomic and molecular systems is dominated by the Coulomb interaction, Eq.~(\ref{eq:Coulomb}), which can be identified by writing the interaction operators in the Coulomb gauge. 

According to Sucher's arguments \cite{sucherPhD1958,DoKr74} (following the field theoretical derivation of the BS equation by Gell--Mann and Low \cite{GMLo51}), it is valid to perform renormalization of the radiative terms in the Feynman gauge, and then, use the resulting expressions for the interacting problem, written in the Coulomb gauge. This special procedure is known as the mixed-gauge representation.

\subsection{Coordinate and Fourier transformation: the total and relative time and energy}
Eq.~(\ref{eq:BSint}) can be rewritten for the wave function of a bound state
(\emph{e.g.,} Ch. 6 of \cite{GrReQEDBook09} or Ch. 12 of \cite{GrossQFTBook99}), formally including now also the radiative effects \cite{DoKr74}, as 
\begin{align}
  \Psitot(x_1, x_2) 
  =
  -\iim
  \int \dd y_1 \dd y_2 \dd y'_1 \dd y'_2\ 
    \Konevp(x_1,y_1)
    \Ktwovp(x_2,y_2)
    \barG(y_1,y_2;y'_1,y'_2) 
    \Psitot(y'_1,y'_2) \; ,
  \label{eq:BSintb}
\end{align}
or in short,
\begin{align}
  \Psitot = -\iim \Konevp \Ktwovp \barG \Psitot \; ,
  \label{eq:BSintbmx}
\end{align}
which, using Eq.~(\ref{eq:invpropagrad}),
can be rearranged to (we note the missing imaginary unit in Ref.~\citenum{DoKr74})
\begin{align}
  \Konev^{-1} \Ktwov^{-1} \Psitot
  &=
  -\iim 
  \left[%
    \barG + \iim \Konev^{-1} \Sigtwov + \iim \Ktwov^{-1} \Sigonev - \iim \Sigonev \Sigtwov
  \right] \Psitot
  =-\iim\kernel'\Psitot \; ,
\end{align}
where the full `interaction' kernel, containing also the radiative corrections, was defined as
\begin{align}  
  \kernel' = \barG + \iim \Konev^{-1}\Sigtwov + \iim \Ktwov^{-1}\Sigonev - \iim \Sigonev \Sigtwov \; .
  \label{eq:fullG}
\end{align}
From rearrangement of the operator form of Eq.~(\ref{eq:Hextfield}), 
$\Konev^{-1}=-\iim \beta_1 [\iim \partial / \partial t_1 -H_1]$  and using $\beta_1\beta_1 = \beta_2\beta_2 =1$, we obtain
\begin{align}
  &\left[%
    \iim \pd{}{t_1}-H_1
  \right]
  \left[%
    \iim \pd{}{t_2}-H_2
  \right] \Psi
  =
  \iim \beta_1 \beta_2 \kernel' \Psitot \ ,
  \label{eq:fullwave} 
\end{align}
which is a (space-time) wave equation which accounts for (non-radiative) interactions and radiative corrections on an equal footing.

Since Eq.~(\ref{eq:fullwave}) includes the `own' time for both particles, but the $U$ external interaction (if any) is time independent in our frame of reference, we can write the two-particle wave function as
\begin{align}
  \Psitot(x_1,x_2) 
  =
  \eem^{-\iim ET} \Psi(\bx_1,\bx_2,t) \;  ,
  \label{eq:stwf}
\end{align}
where the average (`total') time and relative time was introduced as
\begin{align}
  T=\frac{1}{2}(t_1+t_2) 
  \quad\text{and}\quad
  t=t_1-t_2 \ .
\end{align}
It is important to note that $E$ is the total energy of the two-particle system and it corresponds to the $T$ total time, Eq.~(\ref{eq:stwf}).
Similarly to $T$ and $t$, we define 
\begin{align}
  T'=\frac{1}{2}(t'_1+t'_2)
  \quad\text{and}\quad
  t'=t'_1-t'_2 \; .
\end{align}
Then, we obtain the following equation for the 
$\Psi(\br_1,\br_2,t)$ space- and relative-time wave function, 
\begin{align}
  &\left[%
    \frac{E}{2} + \iim \pd{}{t} - H_1
  \right]
  \left[%
    \frac{E}{2} - \iim \pd{}{t} - H_2
  \right]  \Psi(\br_1,\br_2,t)
  \nonumber \\
  &\quad\quad\quad\quad =
  \iim \beta_1 \beta_2
  \int \kernelrelt'(\bx_1,\bx_2, t ; \bx'_1,\bx'_2,t')\Psi(\bx'_1,\bx'_2,t')\ \dd\bx'_1 \dd\bx'_2 \dd t' \; 
  \label{eq:BSreltime}
\end{align}
with the interaction kernel depending only on the relative time variables,
\begin{align}
  \kernelrelt'(\bx_1,\bx_2, t ; \bx'_1,\bx'_2,t')
  =
  \int_{-\infty}^{+\infty}
    \kernel'(x_1,x_2;x'_1,x'_2)\ \eem^{\iim E (T-T')}\ \dd T' \; ,
  \label{eq:kernelcoord}
\end{align}
where it is exploited that the external field is time independent, \emph{i.e.,} $\kernel'(x_1,x_2;x'_1,x'_2)$ depends on $T$ and
$T'$ only through the $T'-T$ difference, and $T$ represents only a constant shift for the $T'$ integration variable.

Both Sucher \cite{sucherPhD1958} and Douglas and Kroll \cite{DoKr74} continued the calculation in momentum space, and we follow this line of thought. The $\bx_1,\bx_2$ space coordinates of the two particles and the $t$ relative time are replaced with the $\bp_1,\bp_2$ momenta and the $\epsi$ relative energy. The relative-time and relative-energy wave functions are connected by the seven-dimensional Fourier transformation, 
\begin{align}
  \Psi(\bx_1,\bx_2,t)
  =
  \frac{1}{(2\pi)^\frac{7}{2}}
  \int_{\mathbb{R}^7}
    \eem^{\iim (\bp_1 \bx_1 + \bp_2 \bx_2 -\epsi t)}\ 
    \Psimom(\bp_1,\bp_2,\varepsilon)\
    \dd \bp_1 \dd \bp_2 \dd \epsi \; ,
\end{align}
while the interaction kernel in momentum space is defined as
\begin{align}
  &\mcG'(\bp_1,\bp_2,\epsi,\bp_1',\bp_2',\epsi')
  = \nonumber \\
  &\quad \frac{{\beta_1 \beta_2}}{{(2\pi)^6}} 
  \int \eem^{-\iim [(\bp_1\br_1+\bp_2\br_2 -\epsi t)-(\bp_1'\br_1'+\bp_2'\br_2' -\epsi' t')]}
    \kernelrelt'(\br_1,\br_2,t,\br_1',\br_2',t')\ 
    \dd \br_1 \dd \br_2 \dd t \dd\br_1' \dd \br_2' \dd t' \; ,
\end{align}
and it acts as an integral operator,
\begin{align}
  \mcG' f(\bp_1,\bp_2,\epsi)
  =
  \int
    \mcG'(\bp_1,\bp_2,\epsi,\bp_1',\bp_2',\epsi')\ 
    f(\bp_1',\bp_2',\epsi')\ 
    \dd \bp_1' \dd \bp_2' \frac{\dd\epsi'}{-2\pi\iim} \; .
  \label{eq:integralOp}
\end{align}
Then, Eq.~(\ref{eq:BSreltime}) can be rewritten as
\begin{align}
  \mcF \Psimom(\bp_1,\bp_2,\varepsilon) 
  =
  \mcG' \Psimom(\bp_1,\bp_2,\varepsilon)
  \label{eq:eiveq}
\end{align}
and
\begin{align}
  \mcF =  \mcF_1  \mcF_2
\end{align}
with
\begin{align}
  \mcF_1 &= \frac{E}{2} + \epsi - \mcH_1(\bp_1) \\
  \mcF_2 &= \frac{E}{2} - \epsi - \mcH_2(\bp_2)
\end{align}
and their inverse define the one-particle propagators (for the $E$ total and $\epsi$ relative energy), which will be used in later sections, 
\begin{align}
  S_1(p_1) &= \mcF^{-1}_1 = 
  \left[% 
    \frac{E}{2}+\varepsilon - \mcH_1(\bp_1)
  \right]^{-1}
  \label{eq:apropag}
  \\
  S_2(p_2) &= \mcF^{-1}_2 = 
  \left[% 
    \frac{E}{2}-\varepsilon - \mcH_2(\bp_2)
  \right]^{-1}  
  \label{eq:bpropag}
\end{align}
with the four-vector variables $p_1=(\bp_1,\epsi)$ and $p_2=(\bp_2,-\epsi)$.
$\mcH_1$ (and similarly $\mcH_2$) is the momentum-space form of the one-particle Dirac Hamiltonian, Eq.~(\ref{eq:oneDirac}). 
In this representation, the interaction operators are integral operators, as it was indicated in Eq.~(\ref{eq:integralOp}) for $\mcG'$. $\mcV_1$ (and $\mcV_2$) labels the external-field Coulomb operator. 
For the example of a single nucleus fixed at the origin, Eq.~\eqref{eq:nucleusfield}, 
\cite{sucherPhD1958,DoKr74}
\begin{align}
  \mcV_1 f(\bp_1,\bp_2,\epsi)
  &=
  z_1\frac{Z\alpha}{2\pi^2} 
  \int 
    \frac{1}{\bk^2} f(\bp_1-\bk,\bp_2,\epsi)\ \dd\bk 
    \label{eq:nucfielda} \\
  \mcV_2 f(\bp_1,\bp_2,\epsi)
  &=
  z_2\frac{Z\alpha}{2\pi^2} 
  \int 
    \frac{1}{\bk^2} f(\bp_1,\bp_2+\bk,\epsi) \ \dd\bk
    \label{eq:nucfieldb} \, .
\end{align}

%%%%%%%%%%%%%%%%%%%%%%%%%%%%%%%%%%%%%%%%%%%%%%%%%%%%%%%%%%%%%%%%%%%%%%%%%%%%%%%%%%%
%%%%%%%%%%%%%%%%%%%%%%%%%%%%%%%%%%%%%%%%%%%%%%%%%%%%%%%%%%%%%%%%%%%%%%%%%%%%%%%%%%%
%%%%%%%%%%%%%%%%%%%%%%%%%%%%%%%%%%%%%%%%%%%%%%%%%%%%%%%%%%%%%%%%%%%%%%%%%%%%%%%%%%%
\subsection{Construction of the interaction kernels using energy-momentum translation operators, the instantaneous part of the interaction \label{sec:kernel}}
According to Eq.~(\ref{eq:fullG}), the full interaction kernel contains contributions  both from `inter-particle' interactions and from radiative contributions, 
\begin{align}
  \mcG' = \mcG_\tI + \mcG^\rad \; .
\end{align}
In what follows, we focus on the construction of the $\mcG_\tI$ inter-particle kernel, which is obtained as the sum of $\mcG_\tI^{(j)}$ operators corresponding to \emph{irreducible} `diagrams' \cite{SaBe51,sucherPhD1958}.  

%%%%%%%%%%%%%%%%%%%%%%%%%%%%%%%%%%%%%%%%%%%%
Action of the interaction kernel for a single-photon exchange (written in the Coulomb gauge), Eqs.~\eqref{eq:onephotCoul}--\eqref{eq:transverse},
on some $f(\bp_1,\bp_2,\epsi)$ two-particle function depending also on the $\epsi$ relative energy
can be written in the momentum-space representation as 
\begin{align}
  \mcG_\col f(\bp_1,\bp_2,\epsi)
  &=
  z_1z_2\frac{\alpha}{2\pi^2}
  \int 
    \frac{1}{\bk^2} f(\bp_1-\bk,\bp_2+\bk,\epsi-\omega)\  
    \dd \bk\ \frac{\dd \omega}{-2\pi\iim} \label{eq:coulact} \\
  \mcG_\tra f(\bp_1,\bp_2,\epsi)
  &=
  z_1z_2 \frac{\alpha}{2\pi^2}
  \int 
    \frac{\ta_1^i \ta_2^i}{\omega^2-\bk^2+\iim\Delta} f(\bp_1-\bk,\bp_2+\bk,\epsi-\omega)\  
    \dd \bk\ \frac{\dd \omega}{-2\pi\iim} \; .
    \label{eq:traact}
\end{align}

A more compact operator form of $\mcG_\col$ and $\mcG_\tra$ is obtained by using $k=(\bk,\omega)$ momentum-energy translation operators. The one-particle translation operators are
\begin{align}
  \eta_1(\bk) f(\bp_1,\bp_2,\epsi)
  &=
  f(\bp_1-\bk,\bp_2,\epsi) 
  \label{eq:etaa}\\
  \eta_2(-\bk) f(\bp_1,\bp_2,\epsi)
  &=
  f(\bp_1,\bp_2+\bk,\epsi) \; ,
  \label{eq:etab}
\end{align}
the two-particle translation operator is, 
\begin{align}
  \eta(k) f(\bp_1,\bp_2,\epsi)
  =
  f(\bp_1-\bk,\bp_2+\bk,\epsi-\omega) \; ,
\end{align}
and for later convenience, we also define the notation
\begin{align}
  \eta(k) = 
  \eta(\bk,\omega) = 
  \eta_1(\bk) \eta_2(-\bk) \eta_\epsi(\omega) \; .
\end{align}
Then, the Coulomb and transverse parts of the one-photon exchange, Eqs.~\eqref{eq:coulact} and \eqref{eq:traact}, can be written as
\begin{align}
  \mcG_\col f(\bp_1,\bp_2,\epsi)
  =
  \int 
    \smallkernel_\col \eta(\bk,\omega) f(\bp_1,\bp_2,\epsi)\  
    \dd \bk  \frac{\dd\omega}{-2\pi\iim} 
\end{align}
with
\begin{align}
  \smallkernel_\col(\bk,\omega) 
  = 
  z_1z_2\frac{\alpha}{2\pi^2} \frac{1}{\bk^2}
\end{align}
and
\begin{align}
  \mcG_\tra f(\bp_1,\bp_2,\epsi)
  =
  \int 
  \ta_1^i \ta_2^i
    \smallkernel_\tra \eta(\bk,\omega) f(\bp_1,\bp_2,\epsi)\  
    \dd \bk  \frac{\dd\omega}{-2\pi\iim} 
\end{align}
with
\begin{align}
  \smallkernel_\tra(\bk,\omega)
  =
  z_1z_2 \frac{\alpha}{2\pi^2}
    \frac{1}
  {\omega^2-\bk^2+\iim\Delta} \; .
\end{align}
It is interesting to note that the Coulomb interaction carries only a trivial shift in the relative-energy dependence, and this corresponds to saying that the Coulomb interaction acts through momentum transfer and the interaction is instantaneous.
At the same time, the transverse part has a non-trivial relative-energy dependence, and this is related to the finite propagation speed of the overall interaction (retardation). 
At the same time, the retardation contribution to the transverse part is small and it is convenient to separate the instantaneous part, which is called the Breit interaction, 
\begin{align}
  \mcG_\tB f(\bp_1,\bp_2,\epsi)
  &=
  \int 
    \ta_1^i \ta_2^i
    \kappa_\tB \eta(\bk,\omega) f(\bp_1,\bp_2,\epsi) \  
    \dd \bk\ \frac{\dd \omega}{-2\pi\iim} \; 
  \label{eq:breitint}
\end{align}
with 
\begin{align}
  \smallkernel_\tB(\bk,\omega) 
  = 
  z_1z_2 \frac{\alpha}{2\pi^2} 
  \frac{1}
  {-\bk^2}    \; .
  \label{eq:bract}
\end{align}
The remainder, \emph{i.e.,} difference of the transverse and the Breit interactions, is the retarded part, which we label as
\begin{align}
  \mcG_\tau 
  &=
  \mcG_\tra-\mcG_\Breit \; ,
  \label{eq:tauact}
\end{align}
while the instantaneous contributions (Coulomb--Breit) can be handled `together', 
\begin{align}
  \mcG_\tCB
  &=
  \mcG_\tC + \mcG_\tB
   \; .
  \label{eq:instact}
\end{align}

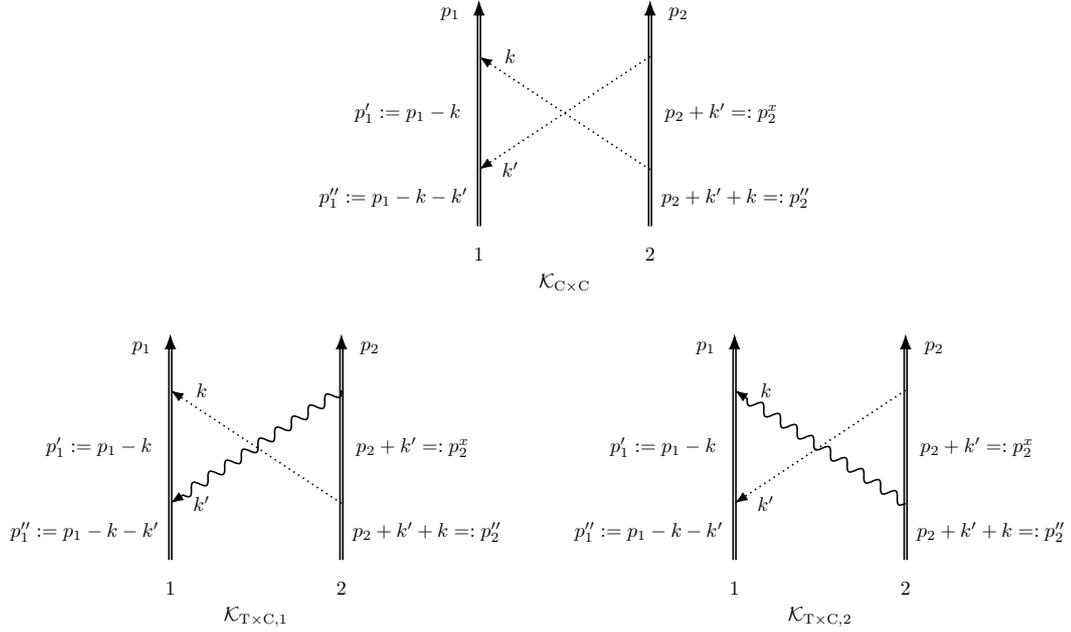
\begin{figure}
  \centering
  \quad\quad
  \scalebox{0.75}{%
    \begin{tikzpicture}
    \draw[boundelectron,-Latex] (0.,0.) -- (0.,4.) ;
    \draw[boundelectron,-Latex] (3.,0.) -- (3.,4.) ;
    \draw[coulomb,thick,-Latex] (3.,3.) -- (0.,1.) ;
    \draw[coulomb,thick,-Latex] (3.,1.) -- (0.,3.) ;
    \node (a0) at (0.,-0.5) {$1$};
    \node (a1) at (-0.5,3.75) {$p_1$};
    \node (a2) at (-1.25,2.0) {$p_1':=p_1-k$};
    \node (a3) at (-1.5,0.5) {$p_1'':=p_1-k-k'$};
    \node (b0) at (3.,-0.5) {$2$};
    \node (b1) at (3.5,3.75) {$p_2$}; 
    \node (b2) at (4.25,2.0) {$p_2+k'=:p_2^x$};
    \node (b3) at (4.50,0.5) {$p_2+k'+k=:p_2''$};    
    \node (c) at (0.55,1.) {$k'$};
    \node (d) at (0.55,3.) {$k$};
    
    \node (CxC) at (1.5,-1.0) {$\mcG_{\tC\times\tC}$};
    \end{tikzpicture} 
    }     
\\~\\
  \scalebox{0.75}{%
    \begin{tikzpicture}
    \draw[boundelectron,-Latex] (0.,0.) -- (0.,4.) ;
    \draw[boundelectron,-Latex] (3.,0.) -- (3.,4.) ;
    \draw[snake=coil, segment aspect=0,thick,-Latex]  (3.,3.) -- (0.,1.) ;
    \draw[coulomb,thick,-Latex] (3.,1.) -- (0.,3.) ;
    \node (a0) at (0.,-0.5) {$1$};
    \node (a1) at (-0.5,3.75) {$p_1$};
    \node (a2) at (-1.25,2.0) {$p_1':=p_1-k$};
    \node (a3) at (-1.5,0.5) {$p_1'':=p_1-k-k'$};
    \node (b0) at (3.,-0.5) {$2$};
    \node (b1) at (3.5,3.75) {$p_2$}; 
    \node (b2) at (4.25,2.0) {$p_2+k'=:p_2^x$};
    \node (b3) at (4.50,0.5) {$p_2+k'+k=:p_2''$};    
    \node (c) at (0.55,1.) {$k'$};
    \node (d) at (0.55,3.) {$k$};
    \node (TxC,a) at (1.5,-1.0) {$\mcG_{\tT\times\tC,1}$};
    \end{tikzpicture} 
    }    
  ~~~~
  \scalebox{0.75}{%
    \begin{tikzpicture}
    \draw[boundelectron,-Latex] (0.,0.) -- (0.,4.) ;
    \draw[boundelectron,-Latex] (3.,0.) -- (3.,4.) ;
    \draw[coulomb,thick,-Latex] (3.,3.) -- (0.,1.);
    \draw[snake=coil, segment aspect=0,thick,-Latex]  (3.,1.) -- (0.,3.) ;
    \node (a0) at (0.,-0.5) {$1$};
    \node (a1) at (-0.5,3.75) {$p_1$};
    \node (a2) at (-1.25,2.0) {$p_1':=p_1-k$};
    \node (a3) at (-1.5,0.5) {$p_1'':=p_1-k-k'$};
    \node (b0) at (3.,-0.5) {$2$};
    \node (b1) at (3.5,3.75) {$p_2$}; 
    \node (b2) at (4.25,2.0) {$p_2+k'=:p_2^x$};
    \node (b3) at (4.50,0.5) {$p_2+k'+k=:p_2''$};    
    \node (c) at (0.55,1.) {$k'$};
    \node (d) at (0.55,3.) {$k$};
    \node (TxC,b) at (1.5,-1.0) {$\mcG_{\tT\times\tC,2}$};
    \end{tikzpicture} 
    }        
    \caption{Example interaction diagrams including four-momentum labels \label{fig:example}}
\end{figure}

To write down the mathematical expression for more complicated $\mcG^{(j)}_\tI$ interactions including multiple (Coulomb and/or transverse) photons (\emph{e.g.,} Fig.~\ref{fig:example}), Sucher \cite{sucherPhD1958} derived and summarized the following simple rules, which we call Sucher's (interaction) rules. \\
First, 
\begin{enumerate}
  \item[(a)]
    label each interaction line with four vectors, $k$, $k'$, $k''$, etc. with assigning each line a specific sense (convenient to choose the same for all lines), \emph{e.g.,} from $2$ to $1$;
  \item[(b)]
    label the final parts of the world lines of the fermions $1$ and $2$ with $p_1=(\bp_1,\epsi)$ and $p_2=(\bp_2,-\epsi)$;
  \item[(c)]
    label all remaining electron lines with conserving the four-momentum.
\end{enumerate}
Second, for a fully labelled diagram, $\mcG^{(j)}_\tI$ can be formulated by writing 
\begin{enumerate}
  \item 
    $\ta^i_1$ or $\ta^i_2$ for a transverse interaction vertex;  
  \item
    a factor $S_1(p_1-k)$ for an intermediate electron line labelled with $p_1-k$ on the path of $1$ and a factor $S_2(p_2+k')$ for an intermediate electron line labelled with $p_2+k'$ on the path of $2$;
    while writing down the factors, it is necessary to preserve the order of events along a world line, \emph{i.e.,} factors for `later' events along a world line stand to the left of factors corresponding to `earlier' events;
  \item
    to the \emph{right} of these expressions a factor $\smallkernel_\tC(k)$ 
    for a Coulomb interaction line labelled with $k=(\bk,\omega)$ and 
    a factor  $\smallkernel_\tT(k')$ for a transverse interaction line with $k'=(\bk',\omega')$;
  \item
    in addition to each $\smallkernel_\tC$ and $\smallkernel_\tT$, 
    an $\eta(k)/(-2\pi \iim)$ factor appears, if the interaction is from $2$ to $1$
    (or, an $\eta(-k)/(-2\pi \iim)$ factor, if the interaction is from $1$ to $2$).  \\
\end{enumerate}

It is also useful to note
that the effect of the $\eta(k)=\eta(\bk,\omega)$ four-momentum translation
on the one-particle propagators, Eqs.~\eqref{eq:apropag}--\eqref{eq:bpropag}, is
\begin{align}
  \eta(k) S_1(p_1) \eta(-k)
  &=
  S_1(p_1-k)
  =
  \left[%
    \frac{E}{2} + \epsi - \omega - \mcH_1(\bp_1-\bk)
  \right]^{-1}
  \label{eq:aproptranslate} \\
  \eta(k) S_2(p_2) \eta(-k) 
  &=
  S_2(p_2+k)
  =
  \left[%
    \frac{E}{2} - \epsi + \omega -\mcH_2(\bp_2+\bk)
  \right]^{-1} \; ,
\end{align}
where the $\bk$-translation of the one-particle Hamiltonians gives 
\begin{align}
  \mcH_1(\bp_1-\bk)
  &=
  \eta_1(\bk) \mcH_1(\bp_1) \eta_1(-\bk)
  =
  \balpha_1 (\bp_1 - \bk) + \beta_1 m + \mcV_1 \\
  \mcH_2(\bp_2+\bk)
  &=
  \eta_2(-\bk) \mcH_2(\bp_2) \eta_2(\bk)
  =
  \balpha_2(\bp_2+\bk) + \beta_2 m + \mcV_2 \; .
\end{align}

Sucher calculated corrections to the energy up to order $\alpha^3\Eh$, Douglas and Kroll calculated the fine-structure splitting up to order $\alpha^4\Eh$, and they have included the following interactions,
\begin{align}
 \mcG_\tI 
 = 
 \mcG_\tC
 +
 \mcG_\tT
 +
 \mcG_{\tC\times\tC} 
 +
 \mcG_{\tT\times\tC}
 +
 \mcG_{\tT\times\tT}  
 +
 \mcG_{\tT\times\tC^2} \; .
\end{align}
It is necessary to compile `by hand' only the irreducible interactions, and all reducible diagrams are automatically included in the solution of the BS equation \cite{SaBe51}.

%%%%%%%%%%%%%%%%%%%%%%%%%%%%%%%%%%%%%%%%%%%%%%%%%%%%%%%%%%%%%%%%%%%%%%%%%%%%%%%%%%%
%%%%%%%%%%%%%%%%%%%%%%%%%%%%%%%%%%%%%%%%%%%%%%%%%%%%%%%%%%%%%%%%%%%%%%%%%%%%%%%%%%%
%%%%%%%%%%%%%%%%%%%%%%%%%%%%%%%%%%%%%%%%%%%%%%%%%%%%%%%%%%%%%%%%%%%%%%%%%%%%%%%%%%%
\subsection{A practical wave equation: the exact equal-time Bethe--Salpeter equation and emergence of the no-pair Dirac--Coulomb(--Breit) Hamiltonian\label{sec:exeqBS}}
Let us exploit the fact that, in atoms and molecules, the dominant part of the interaction is instantaneous (Coulomb or Coulomb--Breit), so, it is convenient to write the kernel as the sum of a $\mcG_\tc$ instantaneous part and the `rest', 
\begin{align}
  \mcG_\tI = \mcG_{\tc} + \mcG_\Delta \; .
\end{align}
The instantaneous part, $\mcG_\tc$, induces only a trivial shift for the $\epsi$ relative energy, and hence the effect of the relative energy can be integrated out 
\begin{align}
  \mcG_\tc \Psimom (\bp_1,\bp_2,\epsi)
  &=
  \int\int 
    \smallkernel_\tc(\bk) \Psimom(\bp_1-\bk,\bp_2+\bk,\epsi-\omega)\ \dd \bk \frac{\dd \omega}{-2\pi \iim}
  \nonumber \\
  &=
  \frac{1}{-2\pi \iim}
  \int \smallkernel_\tc(\bk) \Phi(\bp_1-\bk,\bp_2+\bk)\ \dd \bk \nonumber \\
  &=
  \frac{1}{-2\pi \iim}
  \Ic \Phi(\bp_1,\bp_2)  \; ,
  \label{eq:integrateout}
\end{align}
where $\Ic$ is only a momentum shift integral operator,
\begin{align}
  \Ic \Phi (\bp_1,\bp_2)
  =
  \int \smallkernel_\tc(\bk) \Phi(\bp_1-\bk,\bp_2+\bk)\ \dd\bk \; 
\end{align}
with 
\begin{align}
  \smallkernel_\tc(\bk) 
  = 
  \left\lbrace 
  \begin{array}{@{}cl@{}}
    z_1z_2 \frac{\alpha}{2\pi^2} \frac{1}{\bk^2}\; , & \text{for the Coulomb interaction, $\tc=\tC$} \\
    z_1z_2 \frac{\alpha}{2\pi^2} \frac{1}{\bk^2}(1-\ta_1^i \ta_2^i)\; , & \text{for the Coulomb--Breit interaction,  $\tc=\text{CB}$\; .}  \\
  \end{array}
  \right. \; 
\end{align}
In Eq.~(\ref{eq:integrateout}), it is also important to note 
the emergence of the equal-time wave function, 
\begin{align}
  \Phi(\bp_1,\bp_2) 
  =
  \int_{-\infty}^\infty 
    \Psimom(\bp_1,\bp_2,\epsi)\ \dd \epsi \; .
  \label{eq:equaltimeswf}
\end{align}
Next, we rearrange Eq.~\eqref{eq:eiveq} and separate the instantaneous part of the interaction as 
\begin{align}
  (\mcF-\mcG_\Delta) \Psimom(\bp_1,\bp_2,\epsi)
  &=
  \mcG_{\tc} \Psimom(\bp_1,\bp_2,\epsi)
  \nonumber \\
  \Psimom(\bp_1,\bp_2,\epsi)
  &=
  (\mcF-\mcG_\Delta)^{-1} \mcG_{\tc} \Psimom(\bp_1,\bp_2,\epsi) \; .
  \label{eq:BSrearr1}
\end{align}
By integrating both sides with respect to the relative energy and using Eqs.~\eqref{eq:integrateout} and \eqref{eq:equaltimeswf}, we obtain
\begin{align}
  \int \dd \epsi\ \Psimom
  &=
  \int \dd \epsi\  
  (\mcF-\mcG_\Delta)^{-1}\mcG_{\tc} \Psimom
  \nonumber \\
  \Phi
  &=
  \int \frac{\dd \epsi}{-2\pi\iim}\  
  (\mcF-\mcG_\Delta)^{-1} \Ic \Phi
  \nonumber \\
  \Phi
  &=
  \int \frac{\dd \epsi}{-2\pi\iim}\  
  \mcF^{-1} \Ic  \Phi
  +
  \int \frac{\dd \epsi}{-2\pi\iim}\  
  \mcF^{-1} \mcG_\Delta
  (\mcF-\mcG_\Delta)^{-1} \Ic  \Phi \; ,
  \label{eq:exactequalt1}
\end{align}
where the operator identity was used in the last step, 
\begin{align}
  (A-B)^{-1}
  =
  A^{-1}
  +
  A^{-1} B (A-B)^{-1} \; .
\end{align}
Next, we define the one-particle positive- and negative-energy projection operators for particles $i=1$ and 2 by
\begin{align}
  \mcL_{i\pm}
  =
  \frac{1}{2}
  \left[%
    1 \pm \mcH_i(\bp_i) \mcE_i^{-1}(\bp_i)
  \right] \; ,
  \label{eq:Lproj}
\end{align}
which, at this point, contains a purely formal definition for the one-particle Hamiltonian absolute value operator,
\begin{align}
  \mcE_i(\bp_i) = |\mcH_i(\bp_i)| \; ,
\end{align}
which also means that
\begin{align}
  \mcE_i(\bp_i) \phi^{(i)}_n(\bp_i) 
  = 
  |E^{(i)}_n| \phi^{(i)}_n(\bp_i) \; .
\end{align}
In short, we can also write 
\begin{align}
  \mcH_i \mcL_{i\pm} = \pm \mcE_i \mcL_{i\pm} \;.
  \label{eq:Lproj2}
\end{align}
If there is no external field, \emph{e.g.,} pre-BO description of two spin-1/2 particles, then, $\mcL_{i\pm}$ reduces to the free-particle projector \cite{HaSu84}
\begin{align}
  \mcL_{i\pm}(\bp_i)
  =
  L_{i\pm} (\bp_i)
  = 
  \frac{1}{2}
  \left[%
    1\pm \frac{\balpha_i \bp_i + \beta_i m_i}{\sqrt{p_i^2+m_i^2}}
  \right] 
  \quad (\text{for }U_i=0) \; ,
  \label{eq:Lprojfree}
\end{align}
since for $U_i=0$, $\mcH_i=\mcH_i^{(0)}=\balpha_i\bp_i + \beta_i m_i$ and the eigenvalues of the Hamiltonian absolute value operator are $|E^{(i)}_p|=+\sqrt{p_i^2+m_i^2}$.

Using the $\mcE_1$ and $\mcE_2$ notation, we can write the $\mcF_1^{-1}$ and $\mcF_2^{-1}$ propagators in
$\mcF^{-1}$ as
\begin{align}
  S_1=\mcF_1^{-1}
  =
  \left[%
    \frac{E}{2}+\epsi - \mcH_1 
  \right]^{-1} 
  =
  \frac{\mcL_{1+}}{\frac{E}{2} + \epsi - \mcE_1 + \iim\delta}
  +
  \frac{\mcL_{1-}}{\frac{E}{2} + \epsi + \mcE_1 - \iim\delta} \; ,
  \label{eq:aPropag}
\end{align}
and similarly,
\begin{align}
  S_2=\mcF_2^{-1}
  =
  \left[% 
    \frac{E}{2}-\epsi - \mcH_2
  \right]^{-1} 
  =
  \frac{\mcL_{2+}}{\frac{E}{2} - \epsi - \mcE_2 + \iim\delta}
  +
  \frac{\mcL_{2-}}{\frac{E}{2} - \epsi + \mcE_2 - \iim\delta} \; ,
  \label{eq:bPropag}
\end{align}
according to Feynman's prescription \cite{Fe49a} of adding a complex number with a small negative imaginary value to the mass and the limit is taken from the positive side for a consistent electron-positron theory  (here, the energy replaces Feynman's mass and $\delta>0$ with $\delta \rightarrow 0\plusz$).

The first term in Eq.~(\ref{eq:exactequalt1}) contains a relative-energy integral, but $\Ic$ and $\Phi$ are independent of $\epsi$, so we only need to calculate
\begin{align}
  \int \frac{\dd\epsi}{-2\pi\iim} \mcF^{-1}
  &=
  \int \frac{\dd\epsi}{-2\pi\iim} 
    \mcF_2^{-1}\mcF_1^{-1} \nonumber \\
  &=
  \int \frac{\dd\epsi}{-2\pi\iim}   
    \left(%
      \frac{\mcL_{2+}}{\frac{E}{2} - \epsi - \mcE_2 + \iim\delta}
      +
      \frac{\mcL_{2-}}{\frac{E}{2} - \epsi + \mcE_2 - \iim\delta}
    \right) \nonumber \\
  &\quad\quad\quad\quad\quad
    \left(%
      \frac{\mcL_{1+}}{\frac{E}{2} + \epsi - \mcE_1 + \iim\delta}
      +
      \frac{\mcL_{1-}}{\frac{E}{2} + \epsi + \mcE_1 - \iim\delta} 
    \right) \nonumber \\
  &=
  \int \frac{\dd\epsi}{-2\pi\iim}
    \frac{1}{\frac{E}{2} - \epsi - \mcE_2 + \iim\delta}
    \frac{1}{\frac{E}{2} + \epsi - \mcE_1 + \iim\delta} \mcL_{\pupu} \nonumber \\
  &\ +
  \int \frac{\dd\epsi}{-2\pi\iim}
    \frac{1}{\frac{E}{2} - \epsi - \mcE_2 + \iim\delta}
    \frac{1}{\frac{E}{2} + \epsi + \mcE_1 - \iim\delta}\mcL_{\mipu} \nonumber \\
  &\ +
  \int \frac{\dd\epsi}{-2\pi\iim}
    \frac{1}{\frac{E}{2} - \epsi + \mcE_2 - \iim\delta}
    \frac{1}{\frac{E}{2} + \epsi - \mcE_1 + \iim\delta}\mcL_{\pumi} \nonumber \\
  &\ +
  \int \frac{\dd\epsi}{-2\pi\iim}  
  \frac{1}{\frac{E}{2} - \epsi + \mcE_2 - \iim\delta}
  \frac{1}{\frac{E}{2} + \epsi + \mcE_1 - \iim\delta}\mcL_{\mimi}  \; 
  \label{eq:evalFinv}
\end{align}
with the two-particle projectors, $\mcL_{\sigma \sigma'}=\mcL_{1\sigma} \mcL_{2\sigma'}$ ($\sigma,\sigma'=+$ or $-$). 
%%%%%%%%%%%%%%%%%%%%%%%%%%%%%%%%%%%%%%%%%%%%%%%%%%%%%%%%%%%%%%%%%%%%%%%%%%%%%%%%%%
\par\noindent\rule{\textwidth}{0.4pt}
{\footnotesize%
To evaluate these $\epsi$ integrals, we use Cauchy's residue theorem, 
\begin{align}
   \oint_\gamma f(z) \dd z 
   =
   \text{sgn}(\gamma)\ 2\pi \iim \sum_{k\in\text{poles}} \text{Res}(f,a_k) \; ,
\end{align}
where the summation goes through the poles of $f$ within the domain surrounded by the simple closed curve $\gamma$.
We can choose the positive $\gamma$ contour (O$_\gamma$ counterclockwise, sgn$\gamma=+1$), but identical results are obtained from using the negative $\gamma'$ contour (O$_{\gamma'}$ clockwise, sgn$\gamma'=-1$). Since this is an important step of the calculation, we proceed term by term with the evaluation of Eq.~(\ref{eq:evalFinv}).
\begin{align}
  \lim_{\delta\rightarrow 0\plusz}\mcL_{\pupu} 
  &\int \frac{\dd \epsi}{-2\pi\iim} 
    \underbrace{\frac{-1}{\epsi -\frac{E}{2} + \mcE_2 - \iim\delta}}_{\epsi= \frac{E}{2} - \mcE_2 + \iim\delta \text{\ in O$_\gamma$}}
    \underbrace{\frac{1}{\epsi + \frac{E}{2}  - \mcE_1 + \iim\delta}}_{\epsi=-\frac{E}{2}+\mcE_1-\iim\delta \text{\ in O$_{\gamma'}$}}
    \nonumber \\
  &=
  \left\lbrace 
  \begin{array}{@{}ll@{}}
    \text{for O}_\gamma\ : 
    & 
    \lim_{\delta\rightarrow 0\plusz}    
    \mcL_{\pupu} \frac{2\pi \iim}{-2\pi\iim} \frac{-1}{\frac{E}{2}-\mcE_2+\iim\delta +\frac{E}{2}-\mcE_1 +\iim\delta}
    \\[0.25cm]
    \text{for O}_{\gamma'}\ : 
    & 
    \lim_{\delta\rightarrow 0\plusz}
    \mcL_{\pupu} \frac{-2\pi\iim}{-2\pi\iim} 
    \frac{-1}{-\frac{E}{2}+\mcE_1-\iim\delta -\frac{E}{2}+\mcE_2-\iim\delta} 
  \end{array}
  \right.
  \nonumber  \\
  &=
  \mcL_{\pupu} \frac{1}{E-\mcE_1-\mcE_2}
  \nonumber  \\
  &=
  \mcL_{\pupu} (E-\mcH_1-\mcH_2)^{-1} \\
  \displaystyle\lim_{\delta\rightarrow 0\plusz}\mcL_{\mipu}
  &\int \frac{\dd \epsi}{-2\pi\iim} 
    \underbrace{\frac{-1}{\epsi - \frac{E}{2} + \mcE_2 - \iim\delta}}_{\epsi=\frac{E}{2}-\mcE_2+\iim\delta \text{ in O}_{\gamma}}
    \underbrace{\frac{1}{\epsi + \frac{E}{2} + \mcE_1 - \iim\delta}}_{\epsi=-\frac{E}{2}-\mcE_1+\iim\delta \text{ in O}_{\gamma}}
  \nonumber \\
  &=
  \left\lbrace
  \begin{array}{@{}ll@{}}
    \text{for O}_\gamma: 
    & 
    {\displaystyle\lim_{\delta\rightarrow 0\plusz}}
    \mcL_{\mipu}
    \frac{2\pi\iim}{-2\pi\iim}
    \left[%
      -\frac{1}{\frac{E}{2}-\mcE_2+\iim\delta +\frac{E}{2}+\mcE_1-\iim\delta} 
      +
      \frac{-1}{-\frac{E}{2}-\mcE_1 +\iim\delta -\frac{E}{2}+\mcE_2-\iim\delta}
    \right]
    =
    0
    \\[0.25cm]
    \text{for O}_{\gamma'}: 
    & 
    0 \\
  \end{array}
  \right.
  \nonumber  \\    
  &= 0 \\
  \lim_{\delta\rightarrow 0\plusz}\mcL_{\pumi} 
  &\int \frac{\dd \epsi}{-2\pi\iim} 
    \underbrace{\frac{-1}{\epsi - \frac{E}{2} - \mcE_2 + \iim\delta}}_{\epsi=\frac{E}{2}+\mcE_2-\iim\delta \text{\ in O}_{\gamma'}}
    \underbrace{\frac{1}{\epsi + \frac{E}{2} - \mcE_1 + \iim\delta}}_{\epsi=-\frac{E}{2}+\mcE_1-\iim\delta \text{\ in O}_{\gamma'}}
    \nonumber \\
  &=
  \left\lbrace 
  \begin{array}{@{}ll@{}}
    \text{for O}_\gamma: 
    & 
    0
    \\[0.25cm]
    \text{for O}_{\gamma'}: 
    & 
    {\displaystyle\lim_{\delta\rightarrow 0\plusz}}
    \mcL_{\pumi}
    \frac{2\pi\iim}{-2\pi\iim}
    \left[%
    -\frac{1}{\frac{E}{2}+\mcE_2-\iim\delta + \frac{E}{2}-\mcE_1+\iim\delta}
    +
    \frac{-1}{-\frac{E}{2}+\mcE_1-\iim\delta - \frac{E}{2}-\mcE_2+\iim\delta}    
    \right]
    =
    0
  \end{array}
  \right.
  \nonumber \\
  &=0  \\
{\displaystyle\lim_{\delta\rightarrow 0\plusz}}
  \mcL_{\mimi}
  &\int \frac{\dd\epsi}{-2\pi\iim}  
  \underbrace{\frac{-1}{\epsi -\frac{E}{2} - \mcE_2 + \iim\delta}}_{\epsi=\frac{E}{2}+\mcE_2-\iim\delta \text{ in O}_{\gamma'}}
  \underbrace{\frac{1}{\epsi + \frac{E}{2} + \mcE_1 - \iim\delta}}_{\epsi=-\frac{E}{2}-\mcE_1+\iim\delta \text{ in O}_\gamma}
  \nonumber \\
  &=
  \left\lbrace
  \begin{array}{@{}ll@{}}
    \text{for O}_\gamma: 
    & 
    {\displaystyle\lim_{\delta\rightarrow 0\plusz}}
    \mcL_{\mimi}
    \frac{2\pi\iim}{-2\pi\iim}
    \frac{-1}{-\frac{E}{2}-\mcE_1+\iim\delta-\frac{E}{2}-\mcE_2+\iim\delta}
    =
    -\mcL_{\mimi}\frac{1}{E+\mcE_1+\mcE_2}
    \\[0.25cm]
    \text{for O}_{\gamma'}: 
    & 
    {\displaystyle\lim_{\delta\rightarrow 0\plusz}}
    \mcL_{\mimi}
    \frac{2\pi\iim}{-2\pi\iim}
    (-1)^2\frac{1}{\frac{E}{2}+\mcE_2-\iim\delta + \frac{E}{2}+\mcE_1-\iim\delta}
    =
    -\mcL_{\mimi}\frac{1}{E+\mcE_1+\mcE_2}
     \\
  \end{array}
  \right.
  \nonumber  \\    
  &=-\mcL_{\mimi} \frac{1}{E+\mcE_1+\mcE_2} \nonumber \\
  &=-\mcL_{\mimi} (E-\mcH_1-\mcH_2)^{-1}
\end{align}
}
\par\noindent\rule{\textwidth}{0.4pt}
The result of this short calculation can be summarized as operator identities \cite{sucherPhD1958,DoKr74}
\begin{align}
  \int \frac{\dd \epsi}{-2\pi\iim} 
    \frac{1}{\epsi-A+\iim\delta} \frac{1}{\epsi+B-\iim\delta} &= \frac{1}{A+B}  \\
  \int \frac{\dd \epsi}{-2\pi\iim} 
    \frac{1}{\epsi-A+\iim\delta} \frac{1}{\epsi+B+\iim\delta} &= 0\; ,
\end{align}
where the second identity holds in general, the first is valid only for commuting $A$ and $B$ operators.
All in all, we obtain
\begin{align}
  \int \frac{\dd \epsi}{-2\pi\iim} \mcF^{-1}
  =
  (E-\mcH_1-\mcH_2)^{-1}(\mcL_{\pupu}-\mcL_{\mimi})
  =
  \mcD^{-1}(\mcL_{\pupu}-\mcL_{\mimi}) \; ,
  \label{eq:finv}
\end{align}
where the short notation is introduced,
\begin{align}
  \mcD=E-\mcH_1-\mcH_2 \;.
  \label{eq:ddef}
\end{align}
Using this result, we can re-write Eq.~(\ref{eq:exactequalt1}), 
\begin{align}
  \Phi
  &=
  (E-\mcH_1-\mcH_2)^{-1}(\mcL_{\pupu}-\mcL_{\mimi}) \Ic \Phi
  +
  \int \frac{\dd \epsi}{-2\pi\iim}\  
  \mcF^{-1} \mcG_\Delta
  (\mcF-\mcG_\Delta)^{-1} \Ic \Phi \nonumber \\
  E\Phi
  &=
  [\mcH_1 + \mcH_2 + (\mcL_{\pupu}-\mcL_{\mimi}) \Ic] \Phi
  +
  \mcD
  \int \frac{\dd \epsi}{-2\pi\iim}\  
  \mcF^{-1} \mcG_\Delta
  (\mcF-\mcG_\Delta)^{-1} \Ic \Phi  \; ,
\end{align}
and finally obtain, the exact, equal-time Bethe--Salpeter (eBS) equation \\
\colorbox{white!70!gray}{
\begin{minipage}{1\linewidth}
\begin{align}
  E\Phi
  &=
  (\mcH_1+\mcH_2+\mcL_{\pupu} \Ic\mcL_{\pupu} + \mcH_\Delta) \Phi
  \label{eq:eqtBS}
\end{align}
with 
\begin{align}
  \mcH_\Delta
  =
  \mcL_{\pupu}\Ic(1-\mcL_{\pupu}) - \mcL_{\mimi} \Ic
  +
  \mcD
  \int \frac{\dd \epsi}{-2\pi\iim}\  
  \mcF^{-1} \mcG_\Delta
  (\mcF-\mcG_\Delta)^{-1} \Ic  \; .
  \label{eq:HDelta}
\end{align}
\end{minipage}}
Eq.~(\ref{eq:eqtBS}) is the central equation to our work. It is obtained by equivalent mathematical manipulations from the original space-time Bethe--Salpeter equation, Eq.~\eqref{eq:BSint}, it is a homogeneous, linear equation for the equal-time wave function, $\Phi$, which depends only on the momenta (or coordinates) of the two fermions. At the same time, the exact equal-time equation is a non-linear eigenvalue equation for the $E$ energy, since the $\mcH_\Delta$ term also depends on $E$ (through $\mcF$). 
We can arrive at a useful initial description of atoms and molecules, by first neglecting $\mcH_\Delta$, and starting with the solution of the positive-energy projected or no-pair Dirac--Coulomb(--Breit) equation, 
\begin{align}
  E\Phi
  &=
  (\mcH_1+\mcH_2+\mcL_{\pupu} \Ic \mcL_{\pupu}) \Phi \; .
  \label{eq:npDCB}
\end{align}
It is important to note that in the present derivation \cite{Sa52,sucherPhD1958,DoKr74}, the projector is defined according to Eqs.~(\ref{eq:Lproj}) and (\ref{eq:Lprojfree}), and it is connected to the emergence of the no-pair two-particle Dirac Hamiltonian, Eqs.~\eqref{eq:evalFinv}--\eqref{eq:ddef}.
Variants of the no-pair DC(B) equation are commonly used in relativistic quantum chemistry. 
Sucher \cite{Su83,Su84} analyzed the connection to relativistic quantum chemistry methodologies, in which the Dirac--Hartree--Fock projector is a popular (and natural) choice, and came to the conclusion that the use of that projector is also valid, but then, during the evaluation of the $\mcH_\Delta$ corrections, one has to correct for the difference (which may be complicated). 

During our work, we stick to the original definition, Eq.~(\ref{eq:Lproj}) for two particles in an external field and Eq.~(\ref{eq:Lprojfree}) for an isolated two-fermion system. 
Corresponding numerical results (for helium- and for positronium-like systems) are reviewed in Sec.~\ref{sec:numres}.

During the present calculation, which follows closely the work by Salpeter \cite{Sa52}, Sucher \cite{sucherPhD1958}, and Douglas and Kroll \cite{DoKr74}, it was critical to retain the relative energy between the particles. Integration for the relative energy resulted in the emergence of the no-pair, two-electron Dirac Hamiltonian with instantaneous (Coulomb or Coulomb--Breit) interactions.
Emergence of the two-particle Hamiltonian naturally occurs for a certain choice of the projector. At the moment, we understand $\mcH_\Delta$ in Eq.~\eqref{eq:HDelta} as some `quasi potential' for a DC(B) interacting reference. The DC(B) reference, \emph{i.e.,}
numerical solution of Eq.~(\ref{eq:npDCB}), already contains all reducible interaction diagrams
of the instantaneous kernel \cite{SaBe51}, \emph{i.e.,} the full Coulomb(--Breit) ladder.

%%%%%%%%%%%%%%%%%%%%%%%%%%%%%%%%%%%%%%%%%%%%%%%%%%%%%%%%%%%%%%
\subsection{Phenomenology: why and when the equal-time equation is useful? \label{sec:phenom}}

In atoms and molecules, the interaction of electrons and atomic nuclei (considered now as point-like, quasi-elementary particles) are dominated by electromagnetic forces. 
To capture most of the binding energy in these systems, it is convenient to work in the Coulomb gauge, since the instantaneous Coulomb interaction dominates the binding. 
Subtle magnetic effects can be accounted for by including also the instantaneous Breit interaction in the treatment. 

We can define an equal-time equation, a simple, linear $H\Psi=E\Psi$-type wave equation, by retaining the instantaneous part (Coulomb or Coulomb--Breit) of the interaction mediated by (subsequent exchanges of) a single photon (at a time) and the positive-energy solutions of matter.  The remaining part of the exact equal-time equation can be obtained by integrating through the relative energy (relative time) of the interacting particles (in addition to a simple energy-independent correction term for double-pair instantaneous corrections in the first two terms of Eq.~\eqref{eq:HDelta}). 

The exact equal-time equation form is useful, if the correction obtained from the relative-energy integral is small. In atoms and molecules, it can be anticipated that it is small, because the electromagnetic interaction is relatively weak. 
During the `infinitely' long lifetime of bound systems, infinitely many photon exchanges occur, but these exchanges are mostly consecutive, there are not `many photons' present at the same time. 
The binding of atoms and molecules is dominated by a single photon exchange at a time, and during the lifetime of the system, there is an infinite ladder of single-photon exchanges, the Coulomb ladder or the Coulomb--Breit ladder with non-crossing steps. 
The effect of crossing photons can be identified under a high-energy resolution, and as a small effect it can potentially be accounted for as a (low-order) perturbative correction to the interaction ladder.

If the interaction was much stronger ($\alpha$ was larger), there were more interaction-mediating particles present at the same time, crossed diagrams would be more important, and the equal-time separation and the no-pair approximation would be less useful.

%%%%%%%%%%%%%%%%%%%%%%%%%%%%%%%%%%%%%%%%%%%%%%%%%%%%%%%%%%%%%%%%%%%%%%%%%%%%%%%%%%%%%%%%
%%%%%%%%%%%%%%%%%%%%%%%%%%%%%%%%%%%%%%%%%%%%%%%%%%%%%%%%%%%%%%%%%%%%%%%%%%%%%%%%%%%%%%%%
%%%%%%%%%%%%%%%%%%%%%%%%%%%%%%%%%%%%%%%%%%%%%%%%%%%%%%%%%%%%%%%%%%%%%%%%%%%%%%%%%%%%%%%%
\section{Prospects regarding $\mcH_\Delta$ \label{sec:Hepsi}}
The equal-time two-particle wave equation with instantaneous interactions could be formulated 
at the price of the appearance of a complicated potential energy-like term, which contains an integral with respect to the relative energy of the particles, and which can be considered as some effective potential due to the full-fledged description of the photon field for an interacting two-particle reference (no-pair DC or DCB).

Sucher \cite{sucherPhD1958} formulated low-order perturbative corrections to the no-pair DC(B) energy using
Brillouin--Wigner perturbation theory (BWPT). The advantage of the BWPT energy formula is that it remains formally unchanged for an energy-dependent perturbation (here $\mcH_{\Delta}(E)$),
\begin{align}
  E-E_\tc&
  =
  \langle
    \Phi_\tc| 
    \mcH_{\Delta}
    (1-\Gamma \mcH_{\Delta})^{-1}
    \Phi_\tc\rangle \nonumber \\
    &=
    \langle\Phi_\tc|\mcH_{\Delta}\Phi_\tc\rangle+
           \langle\Phi_\tc|\mcH_{\Delta}\Gamma \mcH_{\Delta}\Phi_\tc\rangle+
           \langle\Phi_\tc|\mcH_{\Delta}\Gamma \mcH_{\Delta}\Gamma \mcH_{\Delta}\Phi_\tc\rangle + ... \ ,
     \label{eq:BWPT}
\end{align}
where the no-pair Hamiltonian is
\begin{align}
  \mcH_\tc = \mcH_1 + \mcH_2 + \mcL_{\pupu} \Ic \mcL_{\pupu}
\end{align}
and $\Gamma(E,\Phi_\tc)$ stands for the (reduced) resolvent
\begin{align}
  \Gamma(E,\phic)
  =
  \sum_n
    \frac{
      | \Phi_{\tc,n} \rangle \langle \Phi_{\tc,n} |
    }{%
      E-E_n
    }
  -
  \frac{
    | \phic \rangle \langle \phic |
  }{%
    E-\Ec
  }  
  =
  (E-\mcH_\tc)^{-1}
  (1-|\phic \rangle \langle \phic | ) \; .
\end{align}
The $\Phi_{\tc,n}$ functions are eigenfunctions of the $\mcH_\tc$ no-pair Hamiltonian with instantaneous (i) interactions.

%%%%%%%%%%%%%%%%%%%%%%%%%%%%%%%%%%%%%%%%%%%%%%%%%%%%%%%%%%%%%%%%%%%%%%%%%%%%%%%%%%
\par\noindent\rule{\textwidth}{0.4pt}
{\footnotesize%
A useful relation for the quasi-Green function is obtained as follows.
Using
the $1 = \mcL_{\pupu} + \mcL_{\pumi} + \mcL_{\mipu} + \mcL_{\mimi}$ completeness relation, 
we can write
\begin{align}
  &\ (E-\mcH_\tc)^{-1} ( 1 - | \phic \rangle \langle \phic | ) \nonumber \\
  &= (E-\mcH_\tc)^{-1} ( \mcL_{\pupu} - | \phic \rangle \langle \phic | )
    +(E-\mcH_\tc)^{-1} ( \mcL_{\pumi} + \mcL_{\mipu} + \mcL_{\mimi} ) \nonumber \\
  &= (E-\mcH_\tc)^{-1} ( \mcL_{\pupu} - | \phic \rangle \langle \phic | )
    +(E-\mcH_1-\mcH_2 - \mcL_{\pupu}\Ic\mcL_{\pupu})^{-1} ( \mcL_{\pumi} + \mcL_{\mipu} + \mcL_{\mimi} ) \nonumber \\
  &= (E-\mcH_\tc)^{-1} ( \mcL_{\pupu} - | \phic \rangle \langle \phic | )
    +(E-\mcH_1-\mcH_2)^{-1} ( \mcL_{\pumi} + \mcL_{\mipu} + \mcL_{\mimi} ) \nonumber \\
  &= (E-\mcH_\tc)^{-1} ( \mcL_{\pupu} - | \phic \rangle \langle \phic | )
    +\mcD^{-1} ( \mcL_{\pumi} + \mcL_{\mipu} + \mcL_{\mimi} )  \; .
\end{align}
Thus, the quasi-Green function can be written as
\begin{align}
  \Gamma(E,\phic)
  =
  (E-\mcH_\tc)^{-1} ( \mcL_{\pupu} - | \phic \rangle \langle \phic | )
    +\mcD^{-1} (1- \mcL_{\pupu})  \; .
\end{align}
}
\par\noindent\rule{\textwidth}{0.4pt}
~\\
%%%%%%%%%%%%%%%%%%%%%%%%%%%%%%%%%%%%%%%%%%%%%%%%%%%%%%%%%%%%%%%%%%%%%%%%%%%%%%%%%%
Aiming for a given order result, simplifications are possible. 
In Sucher's $\alpha^3\Eh$ calculation \cite{sucherPhD1958}, it was sufficient to consider only the first two terms in the expansion of Eq.~\eqref{eq:BWPT}, furthermore, the exact energy could be approximated by $E\approx E_\tc$, \emph{i.e.,} $\mcH_{\Delta}(E)\approx \mcH_{\Delta}(E_\tc)$ and $\Gamma(E,\Phi_\tc)\approx\Gamma(E_\tc,\Phi_\tc)$.
These approximations essentially led to first- and second-order Rayleigh--Schrödinger-type
correction formulae.

For the inclusion of $\mcH_\Delta$ in numerical computations, it is convenient to write it as the sum of two terms,
\begin{align}
  \mcH_\Delta 
  = 
  \mcH_\delta
  +
  \mcH_\epsi \; .
\end{align}
$\mcH_\delta$ is algebraically straightforward, and corresponds to (non-crossing) pair corrections,
\begin{align}
  \mcH_\delta
  =
  \mcL_{\pupu}\Ic (1-\mcL_{\pupu}) - \mcL_{\mimi} \Ic \; .
\end{align}
The technically more involved part includes an integral for the $\epsi$ relative energy 
and carries retardation and crossed-photon contributions (\emph{e.g.,} $\mcG_{\text{C}\times\text{C}}$, $\mcG_{\text{T}\times\text{C},1}$ and $\mcG_{\text{T}\times\text{C},2}$ in Fig.~\ref{fig:example}), 
\begin{align}
  \mcH_\epsi
  =
  \mcD
  \int \frac{\dd \epsi}{-2\pi\iim}\  
  \mcF^{-1} \mcG_\Delta
  (\mcF-\mcG_\Delta)^{-1} \Ic  \; .
\end{align}
For numerical computations, the inverse can be expanded as
\begin{align}
  \mcH_\epsi
  &=
  \mcD
  \int \frac{\dd \epsi}{-2\pi\iim} 
    \mcF^{-1} \mcG_\Delta (\mcF-\mcG_\Delta)^{-1} \Ic
  \nonumber \\
  &=
  \int \frac{\dd \epsi}{-2\pi\iim} 
    \mcD 
    \mcF^{-1} \mcG_\Delta \mcF^{-1} (1-\mcG_\Delta\mcF^{-1})^{-1} \Ic
  \nonumber \\
  &=
  \int \frac{\dd \epsi}{-2\pi\iim} 
    \mcD \mcF^{-1} \mcG_\Delta \mcF^{-1} 
    \left[%
      1
      + \mcG_\Delta\mcF^{-1} + \mcG_\Delta\mcF^{-1} \mcG_\Delta\mcF^{-1} 
      + \ldots
    \right] \Ic
  \nonumber \\
  &=
  \int \frac{\dd \epsi}{-2\pi\iim} 
    \left[%
    \mcD \mcF^{-1} \mcG_\Delta \mcF^{-1} 
    + 
    \mcD \mcF^{-1} \mcG_\Delta \mcF^{-1}\mcG_\Delta\mcF^{-1} 
    +
    \mcD \mcG_\Delta\mcF^{-1} \mcG_\Delta\mcF^{-1} 
    + \ldots
    \right]
    \Ic \; .
\end{align}
We can start by considering the first term of the expansion,
\begin{align}
  \mcH_\epsi^{(1)}
  =
  \mcD \left[\int \frac{\dd \epsi}{-2\pi\iim} 
    \mcF^{-1} \mcG_\Delta \mcF^{-1}\right]  \Ic  \; ,
  \label{eq:nextOrder}
\end{align}
where the $\epsi$ integral is within the square brackets, since it is surrounded by equal-time quantities (without any relative-energy dependence, \emph{cf.} Eqs.~\eqref{eq:eqtBS} and \eqref{eq:HDelta}).
Eq.~(\ref{eq:nextOrder}) can be considered as the next-order term to the known relation, Eq.~(\ref{eq:finv}),
\begin{align}
  \int \frac{\dd \epsi}{-2\pi\iim} 
    \mcF^{-1} 
  =
  \mcD^{-1}(\mcL_{++}-\mcL_{--}) \; .
\end{align}

%%%%%%%%%%%%%%%%%%%%%%%%%%%%%%%%%%%%%%%%%%%%%%%%%%%%%%%%%%%%%%%%%%%%%%%%%%%%%%%%%%
\par\noindent\rule{\textwidth}{0.4pt}
{\footnotesize%
A useful identity for the inverse of the product of the one-particle propagators:
\begin{align}
  \mcF^{-1}
  &=
  \mcF_1^{-1} \mcF_2^{-1}
  \nonumber \\
  &=
  (\mcF_1^{-1}+\mcF_2^{-1}) (\mcF_1+\mcF_2)^{-1} 
  \nonumber \\
  &=
  (\mcF_1^{-1}+\mcF_2^{-1}) \mcD^{-1} 
  =
  \mcD^{-1} (\mcF_1^{-1}+\mcF_2^{-1}) 
  \label{eq:prodOneProp}
  \\
  &=
  (S_1+S_2) \mcD^{-1} 
  =
  \mcD^{-1} (S_1+S_2) \; .
  \label{eq:prodOnePropS}
\end{align}
}
\par\noindent\rule{\textwidth}{0.4pt}
%%%%%%%%%%%%%%%%%%%%%%%%%%%%%%%%%%%%%%%%%%%%%%%%%%%%%%%%%%%%%%%%%%%%%%%%%%%%%%%%%%
Then, using Eq.~\eqref{eq:prodOneProp}, we can write 
\begin{align}
  \mcH^{(1)}_\epsi
  &=
  \mcD\left[\int \frac{\dd \epsi}{-2\pi\iim} 
    \mcF^{-1} \mcG_\Delta \mcF^{-1}\right]  \Ic
  \nonumber \\
  &=
  \mcD\left[\int \frac{\dd \epsi}{-2\pi\iim} 
  \mcD^{-1}(\mcF_1^{-1}+\mcF_2^{-1}) \mcG_\Delta (\mcF_1^{-1}+\mcF_2^{-1}) \right] \mcD^{-1} \Ic
  \nonumber \\
  &=
  \left[\int \frac{\dd \epsi}{-2\pi\iim} 
    (\mcF_1^{-1}+\mcF_2^{-1}) \mcG_\Delta (\mcF_1^{-1}+\mcF_2^{-1}) \right] \mcD^{-1} \Ic
  \nonumber \\
  &=
  \left[\int \frac{\dd \epsi}{-2\pi\iim} 
    (S_1+S_2) \mcG_\Delta (S_1+S_2) \right] \mcD^{-1} \Ic  \; ,
  \label{eq:corr1}
\end{align}
where in the last step, we used the short notation for the one-electron propagators, Eqs.~(\ref{eq:aPropag})--(\ref{eq:bPropag}). It is convenient to consider the propagators as the sum of electronic and positronic contributions, $S_1=S_{1+}+S_{1-}$ and $S_2=S_{2+}+S_{2-}$. 
Even for complicated $\mcG_\Delta$ kernels with multiple photon exchanges, remembering the sign of the imaginary component of the $\epsi$ pole (positive for $S_{1-},S_{2+}$ and negative for $S_{1+},S_{2-}$) is useful for the identification of non-vanishing contributions. 
Furthermore, depending on the actual $\mcG_\Delta$ interaction, one can make arguments (following Sucher \cite{sucherPhD1958} and Douglas \& Kroll \cite{DoKr74}) about the relative importance of the contribution from the  electronic and positronic subspaces.

%%%%%%%%%%%%%%%%%%%%%%%%%%%%%%%%%%%%%%%%%%%%%%%%%%%%%%%%%%%%%%%%%%%%%%%%%%%%%%%%%%
\par\noindent\rule{\textwidth}{0.4pt}
{\footnotesize%
A useful relation regarding $\mcD^{-1}_{\tc} \Ic \Phi_\tc$: \\
If $\Ec$ and $\phic$ is the eigenvalue and eigenfunction of the no-pair Hamiltonian, Eq.~(\ref{eq:npDCB}), then 
\begin{align}
  (\mcH_1+\mcH_2+\mcL_{\pupu} \Ic \mcL_{\pupu}) \phic 
  &= 
  \Ec\phic 
  \nonumber \\
  \mcL_{\pupu} \Ic \mcL_{\pupu} \phic 
  &=
  (\Ec-\mcH_1-\mcH_2)\phic  \nonumber \\
  (\Ec-\mcH_1-\mcH_2)^{-1}\mcL_{\pupu} \Ic \mcL_{\pupu} \phic &= \phic \nonumber \\
  \mcD_{\tc}^{-1}\mcL_{\pupu} \Ic \mcL_{\pupu} \phic &= \phic \nonumber \\
  \mcD_{\tc}^{-1}\mcL_{\pupu} \Ic \phic &= \phic \; ,
  \label{eq:DinvI}
\end{align}
where the last step can be made, since $\phic\in\text{Span}(\mcL_{\pupu})$. This relation with the $\mcD\approx \mcD_{\tc}=E_{\tc}-\mcH_1-\mcH_2$ approximation is used during the calculations.
}
\par\noindent\rule{\textwidth}{0.4pt}
%%%%%%%%%%%%%%%%%%%%%%%%%%%%%%%%%%%%%%%%%%%%%%%%%%%%%%%%%%%%%%%%%%%%%%%%%%%%%%%%%%

To proceed, we can first consider a first-order perturbative correction using the no-pair eigenfunction (of Eq.~\eqref{eq:npDCB}),
\begin{align}
\langle \phic | \mcH^{(1)}_\epsi \phic \rangle 
  &=\langle 
    \phic | 
      \mcD\left[\int \frac{\dd \epsi}{-2\pi\iim} 
      \mcF^{-1} \mcG_\Delta \mcF^{-1}\right]  \Ic
    \phic
  \rangle
  \nonumber \\
  &=
  \langle 
    \phic | 
    \left[\int \frac{\dd \epsi}{-2\pi\iim} 
    (\mcF_1^{-1}+\mcF_2^{-1}) \mcG_\Delta (\mcF_1^{-1}+\mcF_2^{-1}) \right] \mcD^{-1} \Ic
    \phic
  \rangle
  \nonumber \\
  &=
  \langle 
    \phic | 
    \left[\int \frac{\dd \epsi}{-2\pi\iim} 
    (S_1+S_2) \mcG_\Delta (S_1+S_2) \right] \mcD^{-1} (\mcL_{\pupu}+\mcL_{\pumi}+\mcL_{\mipu}+\mcL_{\mimi}) \Ic 
    \phic
  \rangle
  \nonumber \\
  &\approx 
  \langle 
    \phic | 
  \left[\int \frac{\dd \epsi}{-2\pi\iim} 
    (S_1+S_2) \mcG_\Delta (S_1+S_2) \right] \mcD_{\tc}^{-1} \mcL_{\pupu} \Ic 
    \phic
  \rangle  
  \nonumber \\
  &=
  \langle 
    \phic | 
  \left[\int \frac{\dd \epsi}{-2\pi\iim} 
    (S_1+S_2) \mcG_\Delta (S_1+S_2) \right] 
    \phic
  \rangle \nonumber \\
  &=
  \langle 
    \phic | 
    \left[\int \frac{\dd \epsi}{-2\pi\iim} 
    \mcL_{\pupu}
    (S_1+S_2) \mcG_\Delta (S_1+S_2) \mcL_{\pupu} \right] 
    \phic
  \rangle \nonumber \\
  &=
  \langle 
    \phic | 
    \left[\int \frac{\dd \epsi}{-2\pi\iim} 
    (S_{1+}+S_{2+}) \mcG_\Delta (S_{1+}+S_{2+}) \right] 
    \phic
  \rangle \; ,
  \label{eq:approxexpval}
\end{align}
where we used the approximation $\mcD^{-1}\approx \mcD^{-1}_{\tc}=(E_{\tc}-\mcH_1-\mcH_2)^{-1}$, and retained only the positive-energy space contribution between $\mcD^{-1}$ and $\Ic$, furthermore, we inserted the relationship of Eq.~(\ref{eq:DinvI}), 
and exploited the fact that  $\phic$ is the solution of the no-pair equation.

\paragraph{Transverse and retardation correction}
If we consider the solution of the no-pair DC equation, $\phic=\Phi_\tC$ and approximate the total energy in the correction with the no-pair DC energy, $E\approx E_\tC$, then, we can proceed for $\mcG_\Delta=\mcG_\tT$ (Fig.~\ref{fig:example}) by using the fact that the $\epsi$ integral for the `homogeneous' terms ($1+1+$ and $2+2+$) gives zero contributions (the $\epsi$-poles are either both in the positive or in the negative imaginary half plane), and only the mixed ($1+2+$ and $2+1+$) terms have a non-vanishing contribution,
{\footnotesize%
\begin{align}
  \Delta E_{\text{T}\pupu}&=
  \langle 
    \Phi_\tC | 
    \left[\int \frac{\dd \epsi}{-2\pi\iim} 
    (S_{1+}+S_{2+}) \mcG_\tT (S_{1+}+S_{2+}) \right] 
    \Phi_\tC
  \rangle 
  \nonumber \\
  &=
  \langle 
    \Phi_\tC | 
    \left[\int \frac{\dd \epsi}{-2\pi\iim} 
    \lbrace %
      S_{1+}\mcG_\tT S_{2+} 
      +
      S_{2+}\mcG_\tT S_{1+} 
    \rbrace
      \right] 
    \Phi_\tC
  \rangle 
  \nonumber \\
  &=
  \langle 
    \Phi_\tC | 
    \int \frac{\dd \epsi}{-2\pi\iim} 
    \left\lbrace%
    \frac{\mcL_{1+}}{\frac{E_\tC}{2} + \epsi - \mcE_1 + \iim\delta}
    \mcG_\tT
    \frac{\mcL_{2+}}{\frac{E_\tC}{2} - \epsi - \mcE_2 + \iim\delta}
\right.    
\nonumber \\    
&\quad\quad\quad\quad\quad\quad\quad    +
\left.
    \frac{\mcL_{2+}}{\frac{E_\tC}{2} - \epsi - \mcE_2 + \iim\delta}
    \mcG_\tT
    \frac{\mcL_{1+}}{\frac{E_\tC}{2} + \epsi - \mcE_1 + \iim\delta}
    \right\rbrace
    \Phi_\tC
  \rangle \; .
\end{align}
}

Furthermore, $\mcL_{1+}$ and $\mcL_{2+}$ can be suppressed next to the $\Phi_\tC$ 
no-pair wave function (and we assume $z_1z_2=+1$ for simplicity),
{\footnotesize
\begin{align}
  \Delta E_{\text{T}\pupu}^{(12)}
  &=
  \langle 
    \Phi_\tC | 
    \int \frac{\dd \epsi}{-2\pi\iim} 
    \frac{\mcL_{1+}}{\frac{E_\tC}{2} + \epsi - \mcE_1 + \iim\delta}
    \mcG_\tT
    \frac{-\mcL_{2+}}{-\frac{E_\tC}{2} + \epsi + \mcE_2 - \iim\delta}
    \Phi_\tC
  \rangle 
  \nonumber \\
  &=
  \frac{\alpha}{2\pi^2}
  \int\dd\bk \int \frac{\dd\omega}{-2\pi\iim}
  \langle 
    \Phi_\tC | 
    \int \frac{\dd \epsi}{-2\pi\iim} 
    \frac{1}{\epsi + \frac{E_\tC}{2} - \mcE_1 + \iim\delta}
    \frac{\ta_1^i \ta_2^i}{\omega^2-\bk^2+\iim\Delta}
    \eta(\bk,\omega)
    \frac{-1}{\epsi -\frac{E_\tC}{2} + \mcE_2 - \iim\delta}
    \Phi_\tC
  \rangle 
  \nonumber \\
  &=
  \frac{\alpha}{2\pi^2}
  \int\dd\bk \int \frac{\dd\omega}{-2\pi\iim}
  \langle 
    \Phi_\tC | 
    \eta_2(-\bk)
    \ta_2^i 
    \frac{1}{\omega^2-\bk^2+\iim\Delta}
    %\lbrace %
    \Big[%
    \int \frac{\dd \epsi}{-2\pi\iim}   
      \underbrace{%
        \frac{1}{\epsi + \frac{E_\tC}{2} - \mcE_1 + \iim\delta}}_{%
        \epsi=-\frac{E_\tC}{2}+\mcE_1-\iim\delta}
      \frac{-1}{\epsi-\omega - \frac{E_\tC}{2} + \mcE_2 - \iim\delta}
    \Big]
    %\rbrace
    \ta_1^i 
    \eta_1(\bk)
    \Phi_\tC
  \rangle 
  \nonumber \\
  &=
  \frac{\alpha}{2\pi^2}
  \int\dd\bk \int \frac{\dd\omega}{-2\pi\iim}
  \langle 
    \Phi_\tC | 
    \frac{2\pi\iim}{-2\pi\iim} 
    \eta_2(-\bk)
    \ta_2^i 
    \frac{1}{\omega^2-\bk^2+\iim\Delta}
    \frac{-1}{-\frac{E_\tC}{2} + \mcE_1  -\omega - \frac{E_\tC}{2} + \mcE_2 }
    \ta_1^i 
    \eta_1(\bk)
    \Phi_\tC
  \rangle 
  \nonumber \\
  &=
  -\frac{\alpha}{2\pi^2}
  \int\dd\bk \int \frac{\dd\omega}{-2\pi\iim}
  \langle 
    \Phi_\tC | 
    \eta_2(-\bk)
    \ta_2^i 
    \frac{1}{\omega^2-\bk^2+\iim\Delta}
    \frac{1}{E_\tC - \mcE_1 - \mcE_2 +\omega }
    \ta_1^i 
    \eta_1(\bk)
    \Phi_\tC
  \rangle \; .
  \label{eq:calcTcorr1}
\end{align}
}

Then, we proceed along the counter-clockwise integration contour for calculating the $\omega$ integral in Eq.~(\ref{eq:calcTcorr1}), and find one pole at $\omega=-k (+\iim\Delta)$ with $k=|\bk|$, and thus, obtain, 
\begin{align}
  \Delta E_{\text{T}\pupu}^{(12)}
  &=
  -\frac{\alpha}{2\pi^2}
  \int\dd\bk\ \frac{2\pi\iim}{-2\pi\iim}
  \langle 
    \Phi_\tC | 
    \eta_2(-\bk)
    \ta_2^i 
    \frac{1}{2k}
    \frac{1}{E_\tC - \mcE_1 - \mcE_2 - k }
    \ta_1^i 
    \eta_1(\bk)
    \Phi_\tC
  \rangle 
  \nonumber \\
  &=
  \frac{\alpha}{2\pi^2}
  \int\dd\bk\  
  \frac{1}{2k}  
  \langle 
    \Phi_\tC | 
    \eta_2(-\bk)
    \ta_2^i 
    \frac{1}{E_\tC - \mcH_0 - k}
    \ta_1^i 
    \eta_1(\bk)
    \Phi_\tC
  \rangle  \; ,
\end{align}
where $\mcH_0=\mcH_1+\mcH_2$ is the non-interacting two-particle Hamiltonian.
A similar calculation can be carried out for exchanged $1$ and $2$, and thus, the full, positive-energy transverse correction to the no-pair DC wave function is
\begin{align}
  \Delta E_{\text{T}\pupu}
  &=
  \frac{\alpha}{2\pi^2}
  \int\dd\bk\  
  \frac{1}{2k}  
  \langle 
    \Phi_\tC | 
    \left\lbrace
      \ta_2^i     
      \eta_2(-\bk)
      \frac{1}{E_\tC - \mcH_0 - k}
      \eta_1(\bk)
      \ta_1^i   
\right.
      \nonumber \\
&\quad\quad\quad\quad\quad\quad\quad\quad\quad\quad
\left.
      +
      \ta_1^i
      \eta_1(\bk)
      \frac{1}{E_\tC - \mcH_0 - k}
      \eta_2(-\bk)      
      \ta_2^i     
    \right\rbrace
    \Phi_\tC
  \rangle  \; , 
\end{align}
which reproduces Sucher's result, Eq.~(5.26) of Ref.~\cite{sucherPhD1958}. 

With further manipulation, Sucher obtained the Coulomb ladder correction to $T_{++}$ resulting
in the appearance of the interacting no-pair DC Hamiltonian in the resolvent,
\begin{align}
  \Delta E_{\text{T}(\text{C})\pupu}
  =
  \frac{\alpha}{2\pi^2}
  \int\dd\bk\  
  \frac{1}{2k}  
  \langle 
    \Phi_\tC | 
    \left\lbrace
      \ta_1^i         
      \eta_1(\bk)
      \frac{1}{E_\text{C} - \mcH_\tC - k}  
      \eta_2(-\bk)      
      \ta_2^i           
      +
      (1\leftrightarrow 2)     
    \right\rbrace
    \Phi_\tC
  \rangle  \; .
  \label{eq:transversePT}
\end{align}
The important part of this correction is due to the retardation of the interaction, which can be obtained by separating the instantaneous part
according to (p.~75 of  Ref.~\cite{sucherPhD1958})
\begin{align}
  \frac{1}{E_\tC - \mcH_\tC - k}
  =
  -\frac{1}{k}
  +
  \frac{1}{k}
  \frac{E_\tC - \mcH_\tC}{E_\tC - \mcH_\tC - k} \; ,
\end{align}
where the first term gives rise to the Breit operator (\emph{cf.,} Eq.~\eqref{eq:breitint}--\eqref{eq:bract}),
\begin{align}
  B 
  = 
  -\frac{\alpha}{2\pi^2}
  \int \dd\bk\ \frac{1}{k^2}\ta^i_1 \ta^i_2 \eta_1(\bk)\eta_2(-\bk) \; 
  \label{eq:Breit}
\end{align}
and the second term gives the perturbative retardation correction, 
\begin{align}
  \Delta E_\text{ret}
  =
  \frac{\alpha}{2\pi^2}
  \int\dd\bos{k}\ \frac{1}{2k^2}
  \langle %
    \Phi_\tC | 
    \ta^i_1 \eta_1(\bk) 
    \frac{E_\tC - \mcH_\tC}{E_\tC - \mcH_\tC-k}
    \ta^i_2 \eta_2(-\bk)
    \Phi_\tC
  \rangle 
  +(1\leftrightarrow 2) \; .
\end{align}

In 1958, Sucher did not have access to the numerical solution of the no-pair DC equation, so he introduced a series of approximations (including the Pauli approximation) to have final expressions for the non-relativistic wave function.
Nowadays, computer power allows us to compute and converge to `high precision' the numerical solution of the no-pair eigenvalue equation (Sec.~\ref{sec:npDCB}), so it is a challenge to develop algorithms and computational procedures using an accurate relativistic wave function for the evaluation of perturbative corrections of $\mcH_\Delta$. 

For future research, it will be a task to find practical expressions and procedures for the evaluation of the correction terms. Since the correction terms are written in an operator form (without making assumptions about using some special, \emph{e.g.,} one-particle, basis representation), it remains  a technical and computational task to evaluate the integrals for a basis representation allowing high-precision numerical results (Secs.~\ref{sec:npDCB} and \ref{sec:numres}). For general, many-(two-) particle basis functions it may turn out to be convenient to group certain terms together (\emph{e.g.,} retardation and self energy), which would otherwise be evaluated separately (\emph{i.e.,} in computations with one-particle basis functions).

\vspace{0.5cm}
\paragraph{Initial thoughts about crossed photon corrections}
$\mcG_{\tC\times\tC}$ is the simplest crossed-photon correction (Fig.~\ref{fig:example}). Using Sucher's rules (Sec.~\ref{sec:kernel}), we can formulate the correction integral to first-order perturbation theory (and using the $E\approx E_\tc$ approximation),
\begin{align}
  \Delta E_{\tC\times \tC}  
  &=
  \langle 
    \phic | 
    \left[\int \frac{\dd \epsi}{-2\pi\iim} 
    (S_{1+}+S_{2+}) \mcG_{\tC\times \tC} (S_{1+}+S_{2+}) \right] 
    \phic
  \rangle 
  \nonumber \\
  &=
  \int 
    \Phi^\dagger_\tC(\bp_1,\bp_2) 
    \int \frac{\dd \epsi}{-2\pi\iim} \frac{\dd \omega}{-2\pi\iim} \frac{\dd \omega'}{-2\pi\iim}
    [S_{1+}(p_1) + S_{2+}(p_2)]  \nonumber \\
  &\quad\quad\quad\quad\quad\quad\quad\quad
    S_1(p_1-k) S_2(p_2+k') 
    \smallkernel_\tC(\bk) \smallkernel_\tC(\bk')
    \eta(k) \eta(k') \nonumber \\
  &\quad\quad\quad\quad\quad\quad\quad\quad\quad\quad
    [S_{1+}(p_1) + S_{2+}(p_2)] 
    \Phi_\tC(\bp_1,\bp_2)\ 
    \dd\bk\ \dd\bk'\ \dd\bp_1\ \dd\bp_2
  \; ,
\end{align}
which can be simplified by repeated use of the residue theorem \cite{sucherPhD1958}. Direct evaluation (or possible approximation) of the remaining integrals is a future task for precise no-pair wave functions computed by numerical solution of the no-pair wave equation (Secs.~\ref{sec:npDCB}--\ref{sec:numres}).

%%%%%%%%%%%%%%%%%%%%%%%%%%%%%%%%%%%%%%%%%%%%%%%%%%%%%%%%%%%%%%%%%%%%%%%%%%%%%%%%%%%%%%%%
%%%%%%%%%%%%%%%%%%%%%%%%%%%%%%%%%%%%%%%%%%%%%%%%%%%%%%%%%%%%%%%%%%%%%%%%%%%%%%%%%%%%%%%%
%%%%%%%%%%%%%%%%%%%%%%%%%%%%%%%%%%%%%%%%%%%%%%%%%%%%%%%%%%%%%%%%%%%%%%%%%%%%%%%%%%%%%%%%
%
\section{Numerical solution of the no-pair Dirac--Coulomb--Breit eigenvalue equation \label{sec:npDCB}}
This section provides a brief overview of the practical aspects of solving the no-pair Dirac--Coulomb or Dirac--Coulomb--Breit equation with explicitly correlated trial functions \cite{JeFeMa21,JeFeMa22,FeJeMa22,FeJeMa22b}. Explicitly correlated, \emph{i.e.,} two-particle, basis functions make it possible in practice to converge the energy to a precision where comparison of the sixteen-component results with precise and accurate perturbative computations (non-relativistic QED) established in relation with precision spectroscopy is interesting and has been unexplored until recently. For the sake of this comparison, we focus on atoms and molecules of light elements, but in principle, the theoretical and algorithmic framework presented in this work is not limited to low $Z$ systems (unlike finite-order nrQED).

Starting from this section, we replace the natural units ($\hbar=c=\epsilon_0=1$), used in the previous section and common in molecular physics, with Hartree atomic units ($\hbar=e=1/(4\pi\epsilon_0)=m_\text{e}=1$), convenient for quantum chemistry computations. We also note that it is not assumed that the mass of the spin-1/2 particles 
equals the electron mass, and so, we continue to explicitly write out the particle mass.

Furthermore, the practical solution of the no-pair wave equation, Eq.~\eqref{eq:npDCB}, is
carried out in coordinate space, instead of using the momentum-space representation, which was useful for writing down the interactions (Sec.~\ref{sec:kernel}) and working with the propagators (Sec.~\ref{sec:exeqBS}).

\vspace{0.5cm}
\paragraph{No-pair Hamiltonian}
In the coordinate-space representation, the no-pair Dirac--Coulomb--Breit (DCB) Hamiltonian, Eq.~\eqref{eq:npDCB}, is
\begin{align}
 H^{[16]}
 =
 {\cal{L}}_{\pupu}^{[16]}\left(H^{[4]}_1\boxtimes I^{[4]}+I^{[4]}\boxtimes H^{[4]}_2+
 VI^{[16]}
 +B^{[16]}\right){\cal{L}}_{\pupu}^{[16]} \ ,
 \label{eq:nopairDCBH}
\end{align}
where we wrote the projectors around the entire operator, not only around the interaction,
so, we can deal with only the positive-energy block, which was decoupled from the Brown--Ravenhall ($\pumi$ and $\mipu$) and negative-energy ($\mimi$) blocks already in Eq.~\eqref{eq:npDCB}.

$H^{[4]}_i\ (i=1,2)$ is the single-particle Dirac Hamiltonian of
Eq.~(\ref{eq:oneDirac}) shifted by the $m_i c^2$ rest energy,
\begin{align}
 H_i^{[4]}=c(\boldsymbol{\alpha}^{[4]}\cdot\boldsymbol{p}_i)+(\beta^{[4]}-I^{[4]})m_i c^2 +
 U_iI^{[4]} \ ,
\end{align}
and the $U_i$ external potential is due to the nuclei with $Q_A=Z_A$ electric charge, fixed at position~$\bos{R}_A$, 
\begin{align}
  U_i =\sum_{A=1}^{N_\text{nuc}}\frac{z_iZ_A}{|\boldsymbol{r}_i-\boldsymbol{R}_A|} \; ,
  \label{eq:fixedV}
\end{align}
and $z_i$ refers to the electric charge of $i$th active particle. 

The third term of $H^{[16]}$ stands for the Coulomb interaction of the particles
(with $r_{12}=|\bx_1-\bx_2|$)
\begin{align}
  VI^{[16]} = \frac{z_1z_2}{r_{12}}I^{[16]} \; ,
\end{align}
while the last term represents the instantaneous Breit interaction in coordinate representation, \emph{cf.} Eqs.~\eqref{eq:breitint}--\eqref{eq:bract} and \eqref{eq:Breit}:
\begin{align}
 B^{[16]}
 =
 -\frac{z_1z_2}{2r_{12}}
   \left[\boldsymbol{\alpha}^{[4]}\boxtimes\boldsymbol{\alpha}^{[4]}+
                \frac{1}{r_{12}^2}(\boldsymbol{\alpha}^{[4]}\cdot\boldsymbol{r}_{12})
                \boxtimes(\boldsymbol{\alpha}^{[4]}\cdot\boldsymbol{r}_{12})\right] \ .
\end{align}

The symbol $\boxtimes$ stands for a block-wise direct product (also called Tracy--Singh product \cite{TrSi72,LiShLi12,ShLiLi17,SiMaRe15}), which allows us to retain in the many-particle quantities the block structure of the one-particle Dirac matrix expressed with the  $\sigma_i\ (i=1,2,3)$ Pauli matrices, 
\begin{align}
  \alpha^{[4]}_i
  =
  \left(
    \begin{array}{cc}
      0^{[2]} & \sigma_i^{[2]} \\
      \sigma_i^{[2]} & 0^{[2]} \\
    \end{array}
  \right)
  \quad\quad\text{and}\quad\quad
  \beta^{[4]} 
  =
  \left(
    \begin{array}{cc}
      I^{[2]} & 0^{[2]} \\
      0^{[2]} & -I^{[2]} \\
    \end{array}
  \right) \; .
\end{align}
Furthermore, we explicitly indicate the $(k\times k)$  dimensionality of the matrices by the $[k]$ superscript.
For the numerical implementation, we write the Hamiltonian with the $\sigma_i$ Pauli matrices,
\begin{align}
  &H(1,2) = \nonumber \\
  &{ \footnotesize
  \mathcal{L}_{\pupu}^{[16]}
  \left(%
    \begin{array}{@{} c@{}c@{}c@{}c @{}}
       V\unitfour+U \unitfour & 
       c \bsigma\four_2 \cdot \bp_2 & 
       c \bsigma\four_1 \cdot \bp_1 & 
       B\four \\
       c\bsigma\four_2 \cdot \bp_2 & 
       V\unitfour+(U - 2m_2c^2)\unitfour & 
       B\four & 
       c \bsigma\four_1 \cdot \bp_1 \\
       c\bsigma\four_1 \cdot\bp_1 & 
       B\four &
       V\unitfour+(U-2m_1c^2)\unitfour & 
       c \bsigma\four_2 \cdot \bp_2 \\
       B\four & 
       c \bsigma\four_1 \cdot \bp_1 &
       c \bsigma\four_2 \cdot \bp_2 & 
       V\unitfour+(U-2m_{12}c^2)\unitfour \\
    \end{array}
  \right)
  \mathcal{L}_{\pupu}^{[16]}
  } \; ,
  \label{eq:fullHam}
\end{align}
where neglecting the $B$ Breit term (zeroing the anti-diagonal blocks) defines the no-pair Dirac--Coulomb (DC) Hamiltonian,  $H^{[16]}_{\text{DC}}$.

Regarding the $\mathcal{L}_{\pupu}^{[16]}$ projector, it is important to remember that the two-particle Dirac operator with instantaneous (Coulomb or Coulomb--Breit) interactions emerges from the Bethe--Salpeter equation with the ${\cal{L}}_{\pupu}^{[16]}$ operator projecting onto the positive-energy states of the non-interacting problem 
(Sec.~\ref{sec:exeqBS}). 
In this context, a two-particle operator without this projector appears to be an \emph{ad hoc} construct without simple connection to quantum electrodynamics.

\vspace{0.5cm}
\paragraph{Kinetic balance condition and its implementation as a metric}
The no-pair Hamiltonian, Eq.~\eqref{eq:nopairDCBH}, is bounded from below, and thus, development of (precise) variational procedures to solve its eigenvalue equation is highly relevant for practical application of the theory.
To define a good basis set, it is important to ensure a faithful matrix representation of the $\bos{p}^2_i=\bos{p}_i \cdot \bos{p}_i$ identity~\cite{ScWa82}.
During our work, fulfillment of this relation is ensured by using the so-called `restricted kinetic balance' condition, relying on the $(\bos{\sigma}^{[2]}\cdot\bos{a})(\bos{\sigma}^{[2]}\cdot\bos{b})=(\bos{a}\cdot\bos{b})I^{[2]}+\iim (\bos{a}\times\bos{b}) \bos{\sigma}^{[2]}$ property of the Pauli matrices, 
\begin{align}
 X^{[4]}_i=
 \begin{bmatrix}
  I^{[2]} & 0^{[2]} \\
  0^{[2]} & \frac{(\boldsymbol{\sigma}^{[2]}\cdot\boldsymbol{p}_i)}{2m_i c}
 \end{bmatrix}
 \ .
 \end{align}
The simple generalization of this one-particle balance to the two-particle case is
\begin{align}
    X^{[16]}_{12}=X^{[4]}_1\boxtimes X^{[4]}_2=
  \begin{bmatrix}
  I^{[4]} & 0^{[4]} & 0^{[4]} & 0^{[4]} \\
  0^{[4]} & \frac{(\boldsymbol{\sigma}^{[4]}_2\cdot\boldsymbol{p}_2)}{2m_2c} & 0^{[4]} & 0^{[4]} \\
  0^{[4]} & 0^{[4]} & \frac{(\boldsymbol{\sigma}^{[4]}_1\cdot\boldsymbol{p}_1)}{2m_1c} & 0^{[4]} \\
  0^{[4]} & 0^{[4]} & 0^{[4]} & \frac{(\boldsymbol{\sigma}^{[4]}_1\cdot\boldsymbol{p}_1)(\boldsymbol{\sigma}^{[4]}_2\cdot\boldsymbol{p}_2)}{4m_1m_2c^2} 
  \end{bmatrix}
  \ ,
\end{align}
where $\boldsymbol{\sigma}^{[4]}_1=\boldsymbol{\sigma}^{[2]}\otimes I^{[2]}$ and $\boldsymbol{\sigma}^{[4]}_2=I^{[2]}\otimes\boldsymbol{\sigma}^{[2]}$
and $\otimes$ denotes the usual Kronecker product.

We have implemented the kinetic balance condition in an operator form\emph{, i.e.,} as a `metric',  \cite{Ku84,SiMaRe15,JeFeMa21,JeFeMa22,FeJeMa22,FeJeMa22b}
\begin{align}
  H^{[16]}_\text{KB} = X_{12}^{[16]\dagger} H^{[16]} X_{12}^{[16]} 
  \quad \ \text{and} \quad \
  I^{[16]}_\text{KB} = X_{12}^{[16]\dagger} X_{12}^{[16]} \ ,
\end{align}
and detailed operator expressions can be found in Refs.~\cite{JeFeMa21,JeFeMa22,FeJeMa22,FeJeMa22b}.

\vspace{0.5cm}
\paragraph{Spatial and spinor basis}
Finding the spectrum of $H^{[16]}$ (in the $X$-KB metric) numerically requires a finite set of basis functions, which we first define as the product of a (two-particle) spatial function and an elementary spinor (vector), 
\begin{align}
 |\chi^{(16)}_{iq}\rangle
 =
 \varphi_i 
 |e_q^{(16)}\rangle \ .
\end{align}
Regarding the spatial part, two-particle functions can be efficiently represented in the floating explicitly correlated Gaussian (ECG) basis \cite{SuVa98,SiMaRe15,Ma19review},
\begin{align}
 \varphi_i(\boldsymbol{r})
 =
 \exp[-(\boldsymbol{r}-\boldsymbol{s}_i)^\text{T}
  (\boldsymbol{A}^{[2]}_i\otimes I^{[3]})(\boldsymbol{r}-\boldsymbol{s}_i)] \; , \quad i=1,...,\nb\; ,
\end{align}
where $\boldsymbol{r}=(\boldsymbol{r}_1,\boldsymbol{r}_2)^{\text{T}}\in{\mathbb{R}}^6$.
The $\boldsymbol{s}_i\in{\mathbb{R}}^6$ shift vector and the positive-definite $\boldsymbol{A}^{[2]}_i$ matrix elements are parameters of each ECG to be determined via variational optimization \emph{(vide infra)}.
The main advantage of working with ECGs lies in the fact that ECG matrix elements of various operators can be
calculated analytically. 

Regarding the spinor part, 
$4^2=16$ elementary spinors can be constructed, they are of the form
\begin{align}
  |e_q^{(16)}\rangle=
  |\lambda(1)\rangle\otimes|\lambda(2)\rangle
  \otimes
  |\sigma_{m}(1)\rangle\otimes|\sigma_m(2)\rangle \; , \quad q=1,\ldots,16 \ ,
  \label{eq:elemspinor}
\end{align}
with $\lambda=l,s$ (corresponding to the large and small components) and $m=\pm1/2$ (corresponding to the spin projection, $s_z$),
\begin{align}
 |l\rangle=
 \begin{bmatrix}
  1 \\
  0
 \end{bmatrix} \ \ , \ \ 
 %|\uparrow\rangle=
 |s\rangle=
 \begin{bmatrix}
  0 \\
  1
 \end{bmatrix}
 \quad\quad  \text{and} \quad\quad
 |\sigma_{+\frac{1}{2}}\rangle=
 \begin{bmatrix}
  1 \\
  0
 \end{bmatrix} \ \ , \ \ 
 %|\downarrow\rangle=
 |\sigma_{-\frac{1}{2}}\rangle= 
 \begin{bmatrix}
  0 \\
  1
 \end{bmatrix}
 \ .
\end{align}
In Eq.~\eqref{eq:elemspinor}, the `(1)' and `(2)' symbols are shown to highlight the particle index, which is defined by the position of the vector in the Kronecker product.

Instead of the elementary spin representation, we can use a spinor basis which is adapted to the two-particle spin eigenstates ($S=0,M_S=0$ singlet and $S=1,M_S=0,+1,-1$ triplet), \emph{i.e.,}
\begin{align}
 |\chi^{(16)}_{iq}\rangle
 =
 \varphi_i 
 |\tilde{e}_q^{(16)}\rangle 
\end{align}
with 
\begin{align}
  |\tilde{e}_q^{(16)}\rangle
  =
  |\lambda(1)\rangle\otimes|\lambda(2)\rangle
  \otimes
  |\Sigma_{S,M_S}(1,2)\rangle \; 
  , \quad q=1,\ldots,16 \ ,
\end{align}
and
\begin{align}
  |\Sigma_{0,0}\rangle = 
  \frac{1}{\sqrt{2}}
  \begin{bmatrix}
    0 \\
    1 \\
    -1 \\
    0
  \end{bmatrix}  \ ,
\quad
  |\Sigma_{1,0}\rangle = 
  \frac{1}{\sqrt{2}}
  \begin{bmatrix}
    0 \\
    1 \\
    1 \\
    0
  \end{bmatrix}  \ , 
\quad
  |\Sigma_{1,+1}\rangle = 
  \begin{bmatrix}
    1 \\
    0 \\
    0 \\
    0
  \end{bmatrix}  \ , 
\quad
  |\Sigma_{1,-1}\rangle = 
  \begin{bmatrix}
    0 \\
    0 \\
    0 \\
    1
  \end{bmatrix}  \ .
\end{align}
Spin-adapted functions are useful both from an interpretational and also from a practical point of view, as they make a direct connection with non-relativistic results.
This direct connection to non-relativistic computations can be exploited for systems in which relativistic effects are small and the non-relativistic basis parameterization provides a good starting point for relativistic computations \cite{JeFeMa21,JeFeMa22,FeJeMa22,FeJeMa22b,JeMa22}.

\vspace{0.5cm}
\paragraph{Symmetry-adaptation of the basis}
For identical spin-1/2 fermions, anti-symmetrized basis functions must be used,
\begin{align}
  |\psi^{(16)}_{iq}\rangle={\cal{A}}^{[16]}|\chi^{(16)}_{iq}\rangle \ ,
\end{align}
where the ${\cal{A}}^{[16]}$ anti-symmetrizer acts both on the coordinate and spinor space~ \cite{JeFeMa21,JeFeMa22}, 
\begin{align}
 {\cal{A}}^{[16]}=\frac{1}{2}\left[I^{[16]}-(P^{[4]}\otimes P^{[4]}){\cal{P}}_{12}\right]=\frac{1}{2}
  \begin{bmatrix}
  I^{[4]}-P^{[4]}{\cal{P}}_{12} & 0^{[4]} & 0^{[4]} & 0^{[4]} \\
  0^{[4]} & I^{[4]} & -P^{[4]}{\cal{P}}_{12} & 0^{[4]} \\
  0^{[4]} & -P^{[4]}{\cal{P}}_{12} & I^{[4]} & 0^{[4]} \\
  0^{[4]} & 0^{[4]} & 0^{[4]} & I^{[4]}-P^{[4]}{\cal{P}}_{12}
 \end{bmatrix} 
 \ \; ,
\end{align}
where ${\cal{P}}_{12}$ exchanges coordinate space labels and
\begin{align}
  P^{[4]}=
 \begin{bmatrix}
  1 & 0 & 0 & 0 \\
  0 & 0 & 1 & 0 \\
  0 & 1 & 0 & 0 \\
  0 & 0 & 0 & 1
 \end{bmatrix} \; 
 \end{align}
acts on the spinor components.
In particular, $(P^{[4]}{\cal{P}}_{12})(\boldsymbol{\sigma}_1^{[4]}\cdot \boldsymbol{p}_1)=P^{[4]}(\boldsymbol{\sigma}_1^{[4]}\cdot \boldsymbol{p}_2){\cal{P}}_{12}=(\boldsymbol{\sigma}_2^{[4]}\cdot \boldsymbol{p}_2)(P^{[4]}{\cal{P}}_{12})$. 
In two-particle computations with different spin-1/2 particles, the anti-symmetrization step is, of course, omitted~\cite{FeMa22Ps}.

Furthermore, if the system in consideration possesses additional spatial symmetries carried by elements of point group $G$, then, it is useful to adapt the basis functions also to these symmetries.
A ${P}^{[16]}_G$ operation projecting onto an irreducible representation of $G$ can be realized by accounting for both the spatial and spin part of the problem \cite{JeFeMa22,JeMa22}.

\vspace{0.5cm}
\paragraph{A variational procedure}
We approximate eigenfunctions of $H^{[16]}$ in the $\{|\psi^{(16)}_{iq}\rangle\}$ basis by the linear combination, 
\begin{align}
 |\Psi\rangle=\sum_{i=1}^{N_b}\sum_{q=1}^{16}c_{iq}|\psi^{(16)}_{iq}\rangle \; ,
\end{align}
which results in the generalized eigenvalue equation,
\begin{align}
 \boldsymbol{Hc}=E\boldsymbol{Sc} \ , \label{eigvaleq}
\end{align}
where the Hamiltonian and the overlap matrix elements are calculated as
\begin{align}
 (\boldsymbol{H})_{jp,iq}=\langle\psi^{(16)}_{jp}|H^{[16]}_{\text{KB}}|\psi^{(16)}_{iq}\rangle 
 \quad \ \text{and} \quad \ 
 (\boldsymbol{S})_{jp,iq}=\langle\psi^{(16)}_{jp}|I^{[16]}_{\text{KB}}|\psi^{(16)}_{iq}\rangle \ . \label{Hmatelement}
\end{align}
This is a linear variational problem for the $c_{iq}$ coefficients and a non-linear variational problem for the basis function parameters, $\{\boldsymbol{A}^{[2]}_i,\boldsymbol{s}_i\}$. The coefficients are found by solving Eq.~(\ref{eigvaleq}) with a given set of parameters, and the parameters can be refined by minimization of the energy for a selected eigenstate.
This optimization procedure, along with (analytic) evaluation of ECG integrals has been implemented in the QUANTEN program package \cite{Ma19review,FeMa19HH,FeMa19EF,FeKoMa20,IrJeMaMaRoPo21,JeIrFeMa22,JeFeMa21,JeFeMa22,FeJeMa22,FeJeMa22b,MaFe22nad,FeMa22bethe}. 

For calculating the matrix elements of Eq. (\ref{Hmatelement}), the positive-energy projection of the Hamiltonian must be carried out. The matrix representation of the ${\cal{L}}_{\pupu}^{[16]}$ projector is constructed by using the positive-energy eigenstates of the non-interacting two-particle Hamiltonian, $H^{[4]}_1\boxtimes I^{[4]}+I^{[4]}\boxtimes H^{[4]}_2$, represented as a matrix over the actual basis space \cite{JeFeMa21,JeFeMa22,FeJeMa22,FeJeMa22b,FeMa22Ps}.
Selection of the `positive-energy' two-electron states can be realized approximately by `cutting' the non-interacting spectrum based on some energetic condition \cite{JeFeMa21,JeFeMa22}, or more precisely, by rotating the spectrum to the complex plane via complex rescaling of the electron coordinates. This complex rotation (CR) allows us to distinguish three different `branches' of the non-interacting two-electron system (positive-, Brown--Ravenhall, and negative-energy states), in principle, for any finite rotation angle \cite{ByPeKa08}. In practice, an optimal range for the angle can be found by some numerical experimentation (considering the finite precision arithmetic and the finite basis set size).
For the low-$Z$ end of the helium isoelectronic series, the cutting and the CR approaches resulted in practically identical energies, with a relative difference (much) less than one parts-per-billion (ppb) \cite{JeFeMa21,JeFeMa22}.

\vspace{0.5cm}
\paragraph{Perturbative inclusion of $B^{[16]}$}
Since the energetic contribution of the Breit to Coulomb interaction is small, the $B^{[16]}$ term of the DCB Hamiltonian can be treated as a perturbation to the DC problem, which corresponds to the 
$H^{[16]}=H^{[16]}_{\text{DC}}+{\cal{L}}_{\pupu}^{[16]}B^{[16]}{\cal{L}}_{\pupu}^{[16]}$ partitioning. 
The Rayleigh--Schrödinger-type perturbative corrections to the DC energy (up to first or second order)~\cite{FeJeMa22,FeJeMa22b} are evaluated as
\begin{align}
  E^{\pupu}_{\text{DC} \langle \text{B}\rangle,n}
  =
  E^{\pupu}_{\text{DC$,n$}}
  +
  \langle\Psi_n|B^{[16]}_{\text{KB}}|\Psi_n\rangle \ ,
  \label{eq:BreitPTone}  
\end{align}
and
\begin{align}
  {\cal{P}}_n^{(2)}\{B\}
  =
  E^{\pupu}_{\text{DC$,n$}}
  +
  \langle \Psi_n|B^{[16]}_{\text{KB}}|\Psi_n \rangle
  -
  \sum_{k\neq n}
    \frac{%
      |\langle\Psi_k|B^{[16]}_{\text{KB}}|\Psi_n\rangle|^2
    }{%
      E^{\pupu}_{\text{DC$,k$}}-E^{\pupu}_{\text{DC$,n$}}
    } \ ,
 \label{eq:BreitPTtwo}
\end{align}
where $\{|\Psi_n\rangle\}$ and $E^{\pupu}_{\text{DC$,n$}}$ are eigenfunctions and eigenvalues of the no-pair DC Hamiltonian, and $B^{[16]}_{\text{KB}}=X_{12}^{[16]\dagger}B^{[16]}X_{12}^{[16]}$.
Brown--Ravenhall states do not require any further caution in the perturbative calculations either, since the $\Psi_n$ zeroth-order states are within the positive-energy ($\pupu$) space.

For low nuclear charge numbers (low $Z$), the (second-order) perturbative and variational inclusion 
of the Breit interaction resulted in very small energy differences (on the order of a few ppb relative difference)
\cite{FeJeMa22,FeJeMa22b}, which means that the one- and two-Breit photon exchange dominates the `magnetic part' of the interaction.
For higher values of the nuclear charge, the difference between the two approaches 
is anticipated to be larger (to be explored in later work), as higher-order perturbative corrections become more important.
These effects are automatically included in the variational solution, which, after all, can be thought of as the infinite-order summation of ladder diagrams.

On the other hand, it is interesting to note that higher-order corrections due to the `Coulomb ladder' are significant already beyond $Z=1$ \cite{JeFeMa22} (Sec.~\ref{sub:alphascal}), which indicates the relevance of the development of a variational relativistic procedure.

\vspace{0.5cm}
\paragraph{Extension to a pre-Born--Oppenheimer relativistic framework}
We have originally formulated and implemented the equations for two-electron systems with fixed external charges, \emph{i.e.,} for Born--Oppenheimer-like relativistic computations \cite{JeFeMa21,JeFeMa22,FeJeMa22,FeJeMa22b}. 
Most recently, it became possible to generalize these ideas to two-particle systems without external charges, \emph{i.e.,}
pre-Born--Oppenheimer-like \cite{MaRe12,Ma13,FeMa19EF,SaFeMa22} relativistic computations, by using a center-of-momentum frame, by considering the operators and definition of the projector according to Sec.~\ref{sec:exeqBS}, which results in the emergence of a sixteen-component no-pair DC(B) Hamiltonian for the relative (internal) motion. The formalism, implementation details, and numerical results, tested with respect to available perturbative corrections according to Sec.~\ref{sec:numres}, are reported in Ref.~\cite{FeMa22Ps}.

%%%%%%%%%%%%%%%%%%%%%%%%%%%%%%%%%%%%%%%%%%%%%%%%%%%%%%%%%%%%%%%%%%%%%%%%%%%%%%%%%%%%%%%%
%%%%%%%%%%%%%%%%%%%%%%%%%%%%%%%%%%%%%%%%%%%%%%%%%%%%%%%%%%%%%%%%%%%%%%%%%%%%%%%%%%%%%%%%
%%%%%%%%%%%%%%%%%%%%%%%%%%%%%%%%%%%%%%%%%%%%%%%%%%%%%%%%%%%%%%%%%%%%%%%%%%%%%%%%%%%%%%%%
\section{An overview of numerical results\label{sec:numres}}
Before our work, a few `high-precision' Dirac--Coulomb computations have been reported in the literature for helium-like ions \cite{PaGr90,ByPeKa08,PeByKa06,SiMaRe15}, but the different computational procedures (with slightly different technical and theoretical details) delivered (slightly) different numerical results. 
Direct (and useful) comparison of these results with high-precision atomic experiments was not possible due to other, important missing (\emph{e.g.,} radiative and nuclear recoil) corrections. 

At the same time, many questions and concerns appeared in the literature regarding the Dirac--Coulomb(--Breit) `model' taken as a `starting point' and its use in a variational-type approach \cite{BeSabook57,PeByKa06,PeByKa07,ByPeKa08,EsLeSe08,LeSe10}, 
the role and the correct form of the kinetic balance condition \cite{SiMaRe15,EsLeSe08,LeSe10}, 
the `choice' of a good projector for correlated computations \cite{PaGr90,ByPeKa08,AlKnJeDySa16}.
There had been even more controversy (and fewer solid data or formal result) regarding the inclusion of the Breit interaction in a variational treatment.
Most of the observations have their own right in their own context, but the literature was very fragmented and the proper origins for a Dirac--Coulomb-based variational-type procedure with potential utility for precision spectroscopy had been obscure.
At the same time, it is important to add that (various) Dirac--Coulomb(--Breit) Hamiltonian-based computational procedures have already been successfully used for compounds of heavier elements in relativistic quantum chemistry and in relation with (lower) chemical energy resolution, \emph{e.g.,} \cite{dirac2020,BeDSQuTaSt20}.

We have in mind (high) precision spectroscopy experiments for `calculable' systems, calculable to an in principle `arbitrary' precision, if the fundamental equations are known. So, 
we have had anticipated that a precise solution of some (appropriate) variant of a DC(B)-type wave equation is an important step, but it is at most halfway to the solution of the full problem, \emph{i.e.,} for delivering values for direct comparison with precision spectroscopy experiments. 
For this reason, it was of utmost importance to find good anchors for our work to established results and to the (more) complete theory, \emph{i.e.,} relativistic quantum electrodynamics. 

The primary and essential `anchor' for our work was, of course, the connection to the field theoretic Bethe--Salpeter equation that was reviewed in Sec.~\ref{sec:BSeq}. 
This formal connection clearly defines the form of the operator, the projector, and the (wave) equation which we solve, as well as, in principle, all correction terms due to retardation, pair, and radiative effects.

In addition to this formal `benchmark', it was necessary to establish numerical benchmarks to be able to check `intermediate' numerical results.
Extensive testing of numerical results became possible by finding connections to (part of) the already established perturbative relativistic and QED approach based on a non-relativistic reference. 
This perturbative route, sometimes called non-relativistic QED (nrQED), is currently the state of the art for compounds of light elements, which are `calculable' systems to an almost `arbitrary precision',  and has been extensively tested in relation with precision spectroscopy experiments \cite{HaZhKoKa20,AlGiCoKoSc20,test21,HoBySaEiUbJuMe19,PuKoCzPa19,FeKoMa20,SeJaCaMeScMe20}.
The fundamental limitation of nrQED is connected with the finite-order of the available corrections in $\alpha$ (including also $Z\alpha$), which limits the overall accuracy of the results, and this limitation provided the motivation for the present research program.

\subsection{Variational vs. perturbation theory: perturbative benchmark for the no-pair energies 
through $\alpha$ scaling 
\label{sub:alphascal}}
Using a computer implementation of the algorithmic details summarized in Sec.~\ref{sec:npDCB}, we computed the no-pair DC and DCB energies for a series of two-electron atomic and molecular systems with fixed nuclei \cite{JeFeMa21,JeFeMa22,FeJeMa22,FeJeMa22b}, as well as for two-particle positronium-like systems without external charges~\cite{FeMa22Ps}.
Are these numerical results correct? Do they (with the corresponding wave functions) represent a solid intermediate step for further potential computation of increasingly accurate relativistic QED energies for these systems? Direct comparison with experiment, due to missing corrections carried by $\mcH_\Delta$ (and the nuclear motion for the BO-type computations), is not relevant at the current stage.
Numerical results of (more) `complete' nrQED computations have been extensively tested with respect to experiments, and apart from known (and conjectured) limitations of the nrQED framework, these results provide us current numerical benchmarks.

At the same time, comparison of our variational no-pair Dirac--Coulomb(--Breit) energies with  nrQED is not immediately obvious. In nrQED, the total (electronic) energy is written as the sum of the non-relativistic (nr) energy and correction terms for increasing orders of the $\alpha$ fine-structure constant, 
\begin{align}
  E = 
  \underline{E_\text{nr}} 
  + 
  \alpha^2 \underline{\epsi_2} 
  + 
  \alpha^3 \underline{\epsi_3} 
  + 
  \alpha^4 \underline{\epsi_4} 
  + 
  \alpha^5 {\epsi_5} +\ldots \; .
  \label{eq:nrqed}
\end{align}
The $\epsi_2$ correction has been known as the Breit--Pauli Hamiltonian expectation value basically since Breit's work during 1928--1931~\cite{Br28,Br29,Br30,Br31}, the complete $\epsi_3$ correction was first reported by Araki~\cite{araki57} in 1957 and Sucher~\cite{sucherPhD1958} in 1958, the $\epsi_4$ correction to triplet states of helium was derived by Douglas and Kroll~\cite{DoKr74} in 1974 and also for singlet states by Yelkhovsky \cite{Ye01} (and computations with Korobov \cite{KoYe01}) in 2001 and by Pachucki~\cite{Pa06} in 2006. There are currently ongoing efforts \cite{PaYePa20,PaYePa21} for the computation of the $\epsi_5$ correction to triplet states of helium-like systems.
Furthermore, for comparison with experiment, the effect of the nuclear motion is also accounted for in addition to Eq.~\eqref{eq:nrqed}. A recent review provides an overview of the current status for positronium-like systems~\cite{AdCaPR22}.

At the same time, a precise variational solution of the no-pair DCB equation provides us with the no-pair or positive-energy projected energy to all orders of $\alpha$ (all orders of $Z\alpha$), for which the following $\alpha$ series can be formally written as
\begin{align}
  \underline{E^{\pupu}}
  = 
  E_\text{nr} + \alpha^2 \epsi_2 + \alpha^3 \epsi^{\pupu}_3 + \alpha^4 \epsi^{\pupu}_4 + \alpha^5 \epsi^{\pupu}_5 +\ldots \; .
  \label{eq:enerpp}
\end{align}
In Eqs.~\eqref{eq:nrqed} and \eqref{eq:enerpp}, we underlined the quantities that are primarily computed.

In a variational computation, we obtain $E^{\pupu}$ to a certain numerical precision, and we want to check these computations, for \emph{testing} the (correctness of the result of a complex) implementation and computational work and for gaining more \emph{insight and understanding} to the numbers.
We do not directly have access to the formal $\alpha$ expansion of the no-pair energy (right-hand side of Eq.~\eqref{eq:enerpp}), but by computing $E^{\pupu}(\alpha)$ for a series of slightly varied $\alpha$ values, we can fit an $\alpha$-polynomial to the result~\cite{JeFeMa22,FeJeMa22}. 
Coefficients of this fitted polynomial deliver us values for $\epsi_2, \epsi_3^{\pupu}, \epsi_4^{\pupu}, \ldots$ resulting from (a series of) variational computations, and these values can be directly compared (tested) with respect to the relevant (part of the) nrQED corrections (right-hand side of Eq.~\eqref{eq:nrqed}).

The second-order term in Eq.~\eqref{eq:enerpp}, $\epsi_2$, is the same as in Eq.~\eqref{eq:nrqed}.
Beyond second order, the $\epsi^{\pupu}_n$ term contains only part of the $\epsi_n$ `complete' $n$th order nrQED contribution.
Sucher calculated perturbative corrections to the non-relativistic energy \cite{sucherPhD1958} (in this sense, similar in spirit to nrQED), but fortunately, he reported also the no-(and single- and double-)pair part of the contributions. So, we could easily use his no-pair corrections and compare with our $\alpha^3$-order coefficient from variational results of helium-like ions (and two-electron molecules) \cite{JeFeMa22,FeJeMa22,FeJeMa22b}.
Similar $\alpha^3\Eh$-order results are available for hydrogen- and positronium-like two-particle systems from Fulton and Martin \cite{FuMa54}.

All implementation details and extensive comparison with the perturbative results have been reported in Refs.~\cite{JeFeMa21,JeFeMa22,FeJeMa22,FeJeMa22b},
and in the more recent papers, Ref.~\cite{JeMa22} regarding triplet contributions,
and in Ref.~\cite{FeMa22Ps} about a Dirac relativistic pre-Born--Oppenheimer framework for two-particle systems without external charges. 

In a nutshell, an excellent agreement of the no-pair (BO and pre-BO) variational results \cite{JeFeMa21,JeFeMa22,FeJeMa22,FeJeMa22b,JeMa22,FeMa22Ps} is observed through the $\alpha$ scaling procedure for a series of systems, which represents an important milestone for the development of a computational relativistic QED framework for future use in relation with precision spectroscopy.

%%%%%%%%%%%%%%%%%%%%%%%%%%%%%%%%%%%%%%%%%%%%%%%%%%%%%%%%%%%%%%%%%%%%%%%%%%%%%%%%%%%%%%%%
%%%%%%%%%%%%%%%%%%%%%%%%%%%%%%%%%%%%%%%%%%%%%%%%%%%%%%%%%%%%%%%%%%%%%%%%%%%%%%%%%%%%%%%%
%%%%%%%%%%%%%%%%%%%%%%%%%%%%%%%%%%%%%%%%%%%%%%%%%%%%%%%%%%%%%%%%%%%%%%%%%%%%%%%%%%%%%%%%
\section{Summary of the current status and outlook to future work \label{sec:sum}}
\noindent%
With relevance for testing and development of the fundamental theory of atomic and molecular matter,  a relativistic quantum electrodynamics framework for two-spin-1/2 fermion systems (with or without external fixed nuclei) has been reviewed starting from the field theoretic Bethe--Salpeter (BS) equation.
By exploiting the fact that the dominant part of the interaction (Coulomb or Coulomb--Breit) is instantaneous, it is convenient to re-write the original BS equation to an exact equal-time form, which contains the no-pair Dirac--Coulomb(--Breit) Hamiltonian and a correction term, which carries retardation, pair, and radiative corrections. Since this correction term is anticipated to be small, a perturbative treatment has been considered. 
Initial ideas have been reviewed for such a perturbative treatment assuming that a highly precise approximation to the DC(B) wave function, solution of the no-pair equation including the instantaneous Coulomb(--Breit) interaction ladder, is available.

For computing highly precise approximations to the no-pair DC(B) energy and wave function, 
implementation details have been reviewed for an explicitly correlated, variational, no-pair DC(B) computational procedure with the Born--Oppenheimer approximation as well as for extension to a pre-Born--Oppenheimer relativistic framework. 
The computed variational no-pair energies are tested through their $\alpha$ fine-structure constant dependence with respect to the relevant parts of the order-by-order computed non-relativistic QED (nrQED) corrections.

Regarding future work, it is important to realize and implement the evaluation of perturbative corrections for the retardation, pair, and radiative corrections using the variational no-pair DC(B) wave functions already computed for a series of two-particle systems.

Further important work will include generalization of the theoretical framework to $N$-particle systems (including electron, positron, muon, and spin-1/2 nuclei), \emph{i.e.,} which appears to be feasible through the following steps:
(a) starting from an $N$-particle Bethe--Salpeter wave equation;
(b) identification of the relevant irreducible interaction kernels; 
(c) exploitation of the instantaneous character of the dominant part of the interaction;
(d) emergence of the $N$-particle no-pair DCB wave equation for the non-interacting projectors plus a correction term including integral(s) for the relative energy (energies);
(e) solution of the no-pair DCB wave equation to high precision using explicitly correlated basis functions and a variational procedure;
(f) accounting for the retardation, pair, and radiative corrections by perturbation theory;
(g) testing the intermediate results with respect to the relevant terms (if known) from nrQED;
(+$\diamondsuit$)~accounting for the motion of the nuclei (for systems with spin-1/2-nuclei, \emph{e.g.,} H$_2^+$, H$_2$ or H$_3^+$), by using a many-particle pre-Born--Oppenheimer no-pair DCB approach through generalization of Ref.~\citenum{FeMa22Ps}.
At the moment, this endeavour appears to define an extensive research program. The present work reviewed a promising starting point based on the fundamental theory (QED) and outlined necessary practical steps.
Various technical and conceptual difficulties can be foreseen. 

Nevertheless, the success of many-particle Dirac--Coulomb(--Breit) methodologies in relativistic quantum chemistry targeting a much lower, \emph{i.e.,} chemical energy resolution, as well as the limitations due to finite-order nrQED expressions suggest that the development of a computational relativistic QED framework, targeting the spectroscopic energy resolution for testing and further developing the fundamental theory of atomic and molecular matter, is relevant.

\vspace{0.5cm}
\begin{acknowledgments}
\noindent Financial support of the European Research Council through a Starting Grant (No.~851421) is gratefully acknowledged.
We thank Dániel Nógrádi and Antal Jakovác for initial discussions.
\end{acknowledgments}

\vspace{0.5cm}
%\bibliography{references}

\end{document}